\documentclass[fleqn,usenatbib]{mnras}

\usepackage{newtxtext,newtxmath}

\usepackage[T1]{fontenc}

\DeclareRobustCommand{\VAN}[3]{#2}
\let\VANthebibliography\thebibliography
\def\thebibliography{\DeclareRobustCommand{\VAN}[3]{##3}\VANthebibliography}

\usepackage{graphicx}	
\usepackage{amsmath}	
\usepackage{subcaption}
\usepackage[dvipsnames]{xcolor,colortbl}

\newcommand{\calE}{\mathcal{E}}

\definecolor{Gray}{gray}{0.85}
\newcolumntype{a}{>{\columncolor{Gray}}c}
\newcolumntype{?}{!{\vrule width 0.5pt}}
\newcommand{\be}{\begin{equation}}
\newcommand{\ee}{\end{equation}}
\newcommand{\bea}{\begin{eqnarray}}
\newcommand{\eea}{\end{eqnarray}}
\newcommand{\Mdot}{$M\textsubscript{\(\odot\)}$}

\def\bse{\begin{subequations}}
\def\ese{\end{subequations}}
\def\eps{\epsilon}
\def\presub{\vspace{.5cm} \noindent}

\title[The Three-Body Problem]{Testing the Flux-based statistical prediction of the Three-Body Problem}

\author[Manwadkar et al.]{
Viraj Manwadkar,$^{1}$\thanks{E-mail: virajmanwadkar@gmail.com}
Barak Kol,$^{2}$\thanks{E-mail: barak.kol@mail.huji.ac.il}
Alessandro A. Trani,$^{3,4}$
Nathan W. C. Leigh, $^{5,6}$
\\
$^{1}$Department of Astronomy and Astrophysics, University of Chicago, 5640 S. Ellis Ave., Chicago, IL 60637, USA\\
$^{2}$Racah Institute of Physics, Hebrew University, Jerusalem 9190401, Israel\\
$^{3}$Department of Earth Science and Astronomy, College of Arts and Sciences, The University of Tokyo, 3-8-1 Komaba, Meguro-ku, Tokyo 153-8902, Japan \\
$^{4}$Okinawa Institute of Science and Technology, 1919-1 Tancha, Onna-son, Okinawa 904-0495, Japan \\
$^{5}$Departamento de Astronomía, Facultad Ciencias Físicas y Matemáticas, Universidad de Concepción, Concepción 4030000, Chile \\
$^{6}$Department of Astrophysics, American Museum of Natural History, Central Park West at 79th Street, New York, NY 10024, USA \\
}

\date{Accepted XXX. Received YYY; in original form ZZZ}

\pubyear{2021}

\begin{document}
\label{firstpage}
\pagerange{\pageref{firstpage}--\pageref{lastpage}}
\maketitle

\begin{abstract}
We present an extensive comparison between the statistical properties of non-hierarchical three-body systems and the corresponding three-body theoretical predictions. We perform and analyze 1 million realizations for each different initial condition considering equal and unequal mass three-body systems to provide high accuracy statistics. We measure 4 quantities characterizing the statistical distribution of ergodic disintegrations: escape probability of each body, the characteristic exponent for escapes by a narrow margin, predicted absorptivity as a function of binary energy and binary angular momentum, and, finally, the lifetime distribution. The escape probabilities are shown to be in agreement down to the 1\% level with the emissivity-blind, flux-based theoretical prediction. This represents a leap in accuracy compared to previous three-body statistical theories. The characteristic exponent at the threshold for marginally unbound escapes is an emissivity-independent flux-based prediction, and the measured values are found to agree well with the prediction. We interpret both tests as strong evidence for the flux-based three-body statistical formalism. The predicted absorptivity and lifetime distributions are measured to enable future tests of statistical theories.
\end{abstract}




\section{Introduction}
\label{intro}

The general three-body problem is concerned with three masses moving under the influence of their mutual gravitational forces. Some two centuries after this problem was first studied by Newton, \citet{poincare92} discovered chaos in this system, namely, that small perturbations exponentially diverge in time. Therefore, a deterministic mapping from the initial conditions to the final outcomes or an analytical, closed-formed solution for the motion of the three masses are believed to be impossible in the chaotic limit. Using perturbation theory, it is possible to solve for the motion of three-body systems that are hierarchical, i.e., when a well-defined binary pair has formed, and the third body moves on a larger, almost-Keplerian orbit away from the binary. Such hierarchical three-body systems are often stable (e.g. planetary systems). However, finding stable, non-hierarchical three-body systems is not always trivial, especially in dense cluster environments where the regime of stability is very small for compact triples, favoring the formation of triples with long-period outer orbits, which then have short interaction times and tend to be rapidly destroyed \citep[e.g.][]{moore93,leigh16}. In the field and young open clusters, however, the triple fractions can reach of order 10\% \citep{raghavan10}. Unstable, non-hierarchical three-body systems will chaotically disrupt after a three-body close encounter, either into a binary+single or into three individual masses, depending on the total energy of the system.  As a consequence of the chaotic nature, one turns to a \textit{statistical} description, rather than an \textit{analytical} description of the three-body problem. In a statistical description, one aims to obtain the probability distribution of the three-body end-states for an ensemble defined by the mass parameters $m_1,m_2,m_3$ and the conserved quantities: the total energy $E$, the total angular momentum $\vec{L}$ and the total linear momentum $\vec{P}$. Statistics of interest include the escape probability of each mass, the decay time distribution, and the outcome distribution over variables such as the final binary angular momentum. 

Obtaining an accurate, complete statistical theory of the general three-body problem is essential for understanding realistic astrophysical three-body systems like binary-single scattering events in dense star clusters \citep[e.g.][]{hut83,leigh16,leigh18}. Furthermore, such a theory is useful to understand the dynamical interactions of three-body black hole systems and their corresponding final states as one expects to find them in realistic systems like merging galaxies \citep[e.g.][]{loren07} or globular clusters and active galactic nuclei \citep[e.g.][]{samsing18, samsing18_gw}. Gravitational-wave detectors revealed a population of tight binaries consisting of compact objects. Currently, the mechanism behind tightening is a mystery, and one of the leading explanation involves three-body interactions. In this context, an accurate statistical theory would be very useful. Furthermore, the three-body system is not only the historical origin of chaos, but also, it continues to serve as a prototypical chaotic system. The development of a successful statistical theory for the three-body problem will prove to be insightful for other similar systems. 

There have been numerous previous attempts to derive a complete, accurate statistical theory for the three-body problem: \citet{monaghan76a}, \citet{monaghan76b}, \citet{nash78} \citet{Valtonen_book_2006} and more recently \citet{stone19}, \citet{Kol2020} and  \citet{ginat20}. Testing these statistical theories involves running numerical simulations of large ensembles of three-body systems and comparing the final outcome distributions of the simulations to the theoretical predictions. In particular, \citet{Kol2020} introduced a flux-based theory, whose basis differs from all previous statistical theories, and preliminary tests \citet{manwadkar20} have shown it to be highly precise.

This paper's goal is to test a few of the statistical predictions presented in \citet{Kol2020} regarding the three-body outcome distribution and decay time statistics. Section~\ref{theory} reviews the theory, its foundations, and the statistical predictions presented in \citet{Kol2020}. Section~\ref{simulations} describes our numerical three-body simulations and the evolution of general three-body systems to contextualize the discussion. This work is unique in the sense that it does an extensive comparison of three-body simulations with theory over a wide range of three-body mass regimes. Section~\ref{sec:quant_metrics} describes a few quantitative metrics relevant to the three-body system discussion in terms of the time evolution of individual chaotic three-body systems. Section~\ref{thecut} describes the procedure of obtaining the subset of interactions that correspond to ergodic disintegrations. In Section~\ref{results}, we present our key comparisons between three-body simulations and theoretical predictions from \citet{Kol2020}. We present an analytical model for absorptivity in Section~\ref{analyical_model}. The comparisons and results are summarized in Section~\ref{concl}. We also discuss future directions for this work in Section~\ref{future}. 

\section{The Statistical Theory}
\label{theory}

In this section we briefly review the flux-based theory introduced in \citet{Kol2020}. Then we recall and discuss the predictions to be tested in the current paper.

The flux-based theory introduces a statistical theory constructed on a basis which differs from all past statistical treatments starting with \citet{monaghan76a} and up to the review book \citet{Valtonen_book_2006} and the recent closed-form determination of the outcome distribution \cite{stone19}. All previous treatments assume the micro-canonical ensemble, namely, assign probabilities according to phase-space volume. Moreover, they introduce the so-called strong interaction radius in order to guarantee finite phase-space volumes and to exclude an irrelevant part of phase-space (which correspond to causally inaccessible escape scenarios). However, the use of the mirco-canonical ensemble implicitly assumes random (or technically, ergodic) motion throughout phase space, while the system is known to exhibit also regular motion, including the post-decay motion. In addition, setting the value of the strong interaction radius is somewhat arbitrary. The theory of \citet{Kol2020} remedies these flaws. 

Let us start by briefly setting up the problem. The three-body problem can be defined though the Hamiltonian \be
H\left( \{\vec{r}_c,\, \vec{p}_c \}_{c=1}^3 \right) :=  \sum_{c=1}^3 \frac{\vec{p}_c^{~2}}{2 m_c} - \left( \frac{G\, m_1\, m_2}{r_{12}} + cyc. \right)  \label{def:H} 
\ee
where $m_c, ~c=1,2,3$ denote the three masses, $\vec{r}_c$  the bodies' position vectors and $\vec{p}_c$  their momenta, $G$ is Newton's gravitational constant, and $r_{cd}=\left| \vec{r}_c - \vec{r}_d \right|$. 

The conserved charges are the total linear momentum, the total energy, and the total angular momentum denoted by $\vec{P},\, E,\, \vec{L}$, respectively. We work in the center of mass frame.

The conservation laws allow a disintegration of the system. Generally, a non-hierarchical three-body motion ends in this way. Assuming negative total energy, the components of the outgoing states are a binary and a single, also known as the escaper. When they are far apart, the system decouples into a binary subsystem, and an effective hierarchical system defined by replacing the binary by a fictitious object defined by collapsing the binary into its center of mass. The effective system describes the relative motion of the binary and the single. The outcome parameters include the energy and angular momentum of the decoupled binary, denoted by $\eps_B,\, \vec{l}_B$, and those of the effective system, denoted by $\eps_F,\, \vec{l}_F$. 

A special role is played by the binary constant defined by \be
 k := \frac{m_a\, m_b}{m_a + m_b} \left( G m_a m_b\right)^2 \label{def:k}
 \ee
 where $m_a,\, m_b$ are the masses which compose the binary. $k_s$ denotes the binary constant of the binary defined by an escapee $s=1,2$ or $3$.

For a more complete setup,  see \citet{Kol2020}.

The micro-canonical ensemble, used in all previous statistical treatments, suits a closed ergodic system. However, since a three-body state ultimately disintegrates into outgoing components, and by time reversal, the origin of a state is also generically a separated configuration, a more proper context is that of an open system, or a chaotic scattering problem. Therefore, the system displays asymptotic states as well as a chaotic interaction region. In addition to these two kinds of motion, the three-body system allows also for sub-escape excursions, which are a part of motion where the system clearly separates into a binary and a single which fly away from each other, yet their relative velocity is below the escape velocity and hence they are bound to fall back into the interaction region. Accordingly, the sub-escape excursions can be called quasi-asymptotic states.
 
The first part of the flux-based theory considers the system's probability distribution as a time-dependent variable. The probability is discretely distributed between the ergodic region, the sub-escape excursions and the asymptotic states, and the latter two probabilities are further continuously distributed over their parameters. This distribution differs from the micro-canonical ensemble of previous treatments. The time evolution of the distribution is formulated through system of equations (2.35) of \citet{Kol2020}, whose solution describes the statistics of outcome parameters and decay times. 

The second part of the theory involves the differential decay rate out of the ergodic region, which is an essential ingredient in the equation for the above-mentioned statistical evolution of the system. Its distribution over all the asymptotic state parameters $u$ is denoted by $d\Gamma(u)$, and it can be shown to factorize exactly into \be
	d\Gamma_s(u) \propto  \calE (u) \cdot \frac{\sqrt{k_s}\, d\eps_B}{(-2 \eps_B)^{3/2}} \, \frac{1}{l_B\, l_F} d^3l_B\, d^3 l_F\, \delta^{(3)}\left(\vec{l_B}+\vec{l_F}-\vec{L}\right)
\label{dGamma}
\ee
where $s=1,2,3$  is the escaper identity, $\calE$ is the chaotic emissivity ($=$ absorptivity) and $k_s$ is defined in (\ref{def:k}). The variables $\eps_B, l_B$ range over the domain \be
-2\, \eps_B\, l_B^2  \le  k_s ~, \qquad
 \eps_B  \le  E  
\label{binary_region}
\ee

Chaotic absorptivity $\calE(u)$ is defined to be the probability that a scattering of a single off a binary will evolve into a chaotic trajectory, rather than a regular scattering such as a flyby or regular exchange leading to prompt ejection, and the scattering parameters are specified by $u$, see eq. 2.23 of \citet{Kol2020}. In analogy with Kirchhoff's law of thermal radiation, the chaotic absorptivity with incoming parameters $u$ equals chaotic emissivity with outgoing parameters $u$. Clearly, $\calE(u)$ serves to account for the division of phase space into regions of regular and chaotic motion. We shall often omit the adjective ``chaotic'' since we do not discuss other types of emissivity.   

The rest of the expression accounts for the flux of phase space volume into the asymptotic states. This factor reflects the fact that this theory is based on the framework of an open, rather than closed, chaotic system, where the flux of phase-space volume replaces the volume itself as a measure of probability. Hence, it is called the flux-based theory. To have some insight into the form of the flux factor, note that $2 \pi \sqrt{k_s}/(-2 \eps_B)^{3/2}$ is the binary period, and the denominator factors $l_B,\, l_F$ each originate in the central force nature of the respective systems. 

By definition, $\calE(u)$ is bounded to the range $0 \le \calE (u) \le 1$. Otherwise, so far, it is an unknown function of the asymptotic parameters $u$. In this way, (\ref{dGamma}) factors out the outgoing flux and reduces the determination of $d\Gamma(u)$ to that of $\calE(u)$. For derivations, see \citet{Kol2020}.

Altogether, through these ingredients the flux-based theory addresses and improves upon the above-mentioned issues of the previous statistical theories. 

\subsection{Pertinent predictions} 
\label{pert_pred}

We focus on the prediction for the outcome distribution for escapes which originate in the ergodic region, and hence are described by (\ref{dGamma}). Moreover, since $\calE(u)$ is little known, we seek observables which are independent of it, or at least, weakly dependent. In a future work, we intend to test further quantities such as the outcome distribution over binary energy, angular momentum and eccentricity.

\presub {\bf Ergodic escape probability}. To obtain the relative frequency of each escaper mass, one must integrate (marginalize) over all the continuous outcome parameters within the domain (\ref{binary_region}). This process can be said to define a $u$-global observable. Integration averages over the little known $\calE(u)$, and this motivates an approximation where $\calE$ is ignored altogether. As we shall see, the agreement with the numerical simulations is surprisingly good, as long as the masses are comparable to within some measure.

The numerical simulations considered in this paper are set-up such that $L \ge l_{0s}$ where \be
 l_{0s} := \sqrt{k_s/(-2 E)}
\label{def:l0s}
\ee
 is the maximum possible value for the binary angular momentum. For this case one finds that the unnormalized ergodic escape probabilities are given by eq. (2.37) in \citet{Kol2020}, namely, they reduce to the following simple expression  \be
 P_s \propto k_s^{3/2} ~.
\label{escape_prob}
\ee 

The expression was simplified through the well-known fact that the gravitational potential outside a uniform spherical shell can be found by collapsing the shell onto its center, only here this integral is carried out in angular momentum space, the shells have constant $l_B$ and the $1/|\vec{r}-\vec{r}'|$ integrand is analogous to the $1/l_F$ factor. In a second step in the derivation, one first obtains $P_s =l_{0s}^3$, which then implies (\ref{escape_prob}) after a change in normalization. 

\presub {\bf Escape by a narrow margin}. Whereas the escape probability is a global quantity, weakly dependent on $\calE$, here we consider the critical behavior at low $l_F$ which is a $u$-local observable . 

The setup of our numerical simulation is such that conservation of angular momentum implies a minimal or critical value for $l_F$  \be
 l_F \ge l_{F,c} := L - l_{0s}  \ge 0 ~.
 \ee

The form of the $l_F$ probability distribution near this threshold is dominated by the shape of the integration domain (\ref{binary_region}). At threshold, the relevant slice of the domain is empty, and it grows together with the value of $l_F$. Since the energy of the relative motion is close to zero, absorptivity is expected to be high, and in particular, non-zero and smooth.  Since the integrand, namely (\ref{dGamma}), is smooth and hence locally constant, the result is independent of $\calE(u)$ and the integral is dominated by the slice volume, namely the ``phase space factor'', and this leads to a local form of the distribution given by eq. (4.11) of \citet{Kol2020}, namely \be
 dP_s \propto \left( l_F- l_{F,c} \right)_+^2\, dl_F
\label{lF_distrib1}
 \ee
where $(x)_+$ denotes the ramp function, namely $(x)_+=x$ for $x \ge 0$ and $(x)_+=0$ for $x \le 0$. In fact, 

In the language of critical phenomena, the predicted distribution (\ref{lF_distrib1}) has a characteristic exponent 2.

\presub {\bf Predicted absorptivity}. A numeric measurement of the bi-variate outcome distribution $dP_s=dP_s(\eps_B, l_B)$ can be used to predict the value of the  $\calE$, the absorptivity, or equivalently, the emissivity, through (\ref{dGamma}), yielding \be
\bar{\calE}_{\rm s,pred}\left( \eps_B, l_B \right) / \left< \calE _s \right> = \frac{k_s}{6} \frac{(\eps_B/E)^{3/2}}{l_B} \frac{dP_s}{d\eps_B \, dl_B}
 \label{calE_pred}
\ee
where $\bar{\calE}_{\rm s,pred}\left( \eps_B, l_B \right)$ denotes the predicted absorptivity averaged while keeping $\eps_B, l_B$ fixed and $\left< \calE _s \right>$ denotes a global average of absorptivity that is independent of $\eps_B, l_B$. As we are dividing the absorptivity by its global average, we expect the left-hand side (LHS) to have a minimum value of zero and a maximum value larger than 1 but of order 1.  

\presub {\bf Lifetime}. Consider the decay lifetime for a state in the ergodic region. The statistical evolution equations imply that the distribution of lifetimes at early times is given by \be
dP_s = \Gamma_{s,{\rm esc}}\, \exp \left(-\Gamma_{\rm tot}\,  t \right) 
\label{lifetime}
\ee
where $\Gamma_{s,{\rm esc}}$ is the rate of escape into an escape of $s$, and $\Gamma_{\rm tot}$ is the total decay rate into both escapes and sub-escapes. This exponential distribution is a consequence of the ergodic nature of the motion. After some time the distribution becomes dominated by motions that experienced a large sub-escape, and the distribution of decay times crosses over to a 5/3 power-law. Numerical studies of the three-body problem have shown conclusive evidence for this behavior of the lifetime distributions, see e.g. \citet{orlov10}, \citet{manwadkar20}. Derivations of the both the exponential and the 5/3 power-law distributions are given in appendix \ref{appendix:lftm_drv}.

\section{Simulations}
\label{simulations}

\subsection{TSUNAMI N-Body code}
\label{code}

We run the three-body simulations with the regularized N-body code \textsc{tsunami}. \textsc{tsunami} integrates the equations of motion derived from a time-extended Hamiltonian, following the leapfrog algorithm described in \citet{mikkola99a}. Because the basic algorithm is only 2nd order in accuracy, \textsc{tsunami} employs a Bulirsch-Stoer extrapolation scheme \citep{stoer80} that greatly improves the accuracy of the final positions and velocities. Finally, to limit the errors arising from floating point accuracy, \textsc{tsunami} implements a chain coordinate system, wherein particle's coordinates are stored with respect to their closest neighbour, rather than to the center of mass of the system \citep{mikkola1993}. The combination of these techniques make \textsc{tsunami} ideally suited to simulate compact systems of strongly interacting particles, as in the chaotic three-body problem.

\textsc{tsunami} includes additional features, such as perturbative forces  (1PN, 2PN and 2.5PN post-Newtonian corrections, \citealt{blanchet14}; tidal interactions \citealt{hut81,samsing18_tides}) and collision detection; these features are disabled here because we focus on purely Newtonian dynamics of point masses. More details on the code will be presented in a following work (A.A.Trani et. al, in preparation).

To determine the state of the three-body system, we adopt the same classification scheme employed in \citet{manwadkar20}. In brief, this classification scheme is based on energy and stability criteria. The hierarchy state of the triple is checked at every timestep by selecting the most bound pair and checking its relative energy with respect to the third body. If the third body is unbound, we check if the binary-single pair is converging or diverging on a hyperbolic orbit: if diverging, we record the breakup time $\tau_{D}$ and stop the simulation when the single is 10 times the binary semi-major axis away from the binary centre of mass; if otherwise the binary-single is converging, we estimate the time of closest approach and continue the simulation. We also implement criteria to detect stable or meta-stable triples; however these outcomes are forbidden in the context of the parameter space explored in this work.

The code has been modified to track and record the statistics of a three-body interaction, such as the time spent in a hierarchical configuration $\tau_{sub}$ and other useful metrics (see Section~\ref{sec:quant_metrics}).

\subsection{Numerical Setup}
\label{simexp}

\begin{figure}
\centering 
\includegraphics[width=\columnwidth]{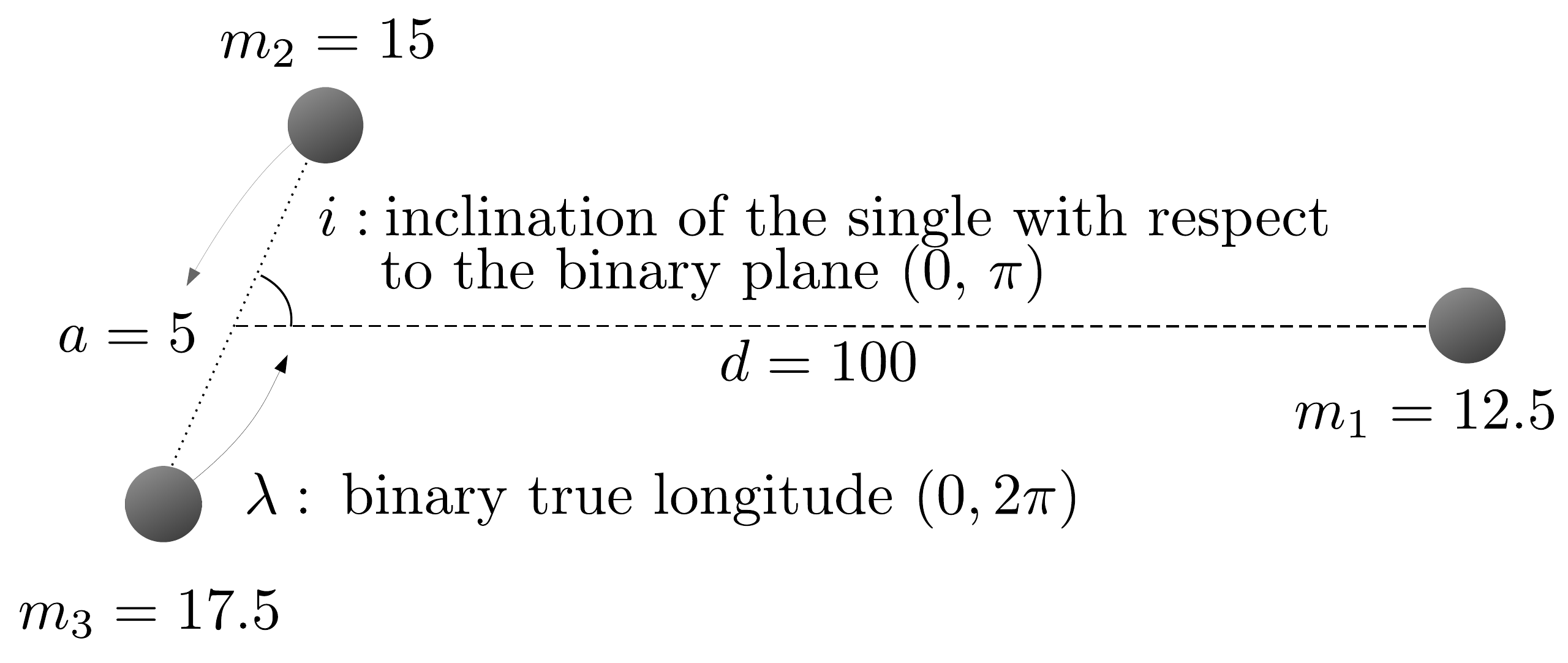}
\caption{Scheme of the initial setup of the simulations. Units are in au and M$_\odot$. The masses corresponds to the second set number in Table~\ref{tab:sims_summary}. All the other parameters are the same for each set.}
\label{fig:icscheme}
\end{figure}

In this work, we consider 9 different three-body systems with different particle masses; each set comprises $10^6$ realizations. The three-body systems range from the equal mass system (15-15-15\Mdot) to unequal mass systems (5-15-25\Mdot). The initial setup is that the heaviest two masses form a binary pair moving in a circular orbit and the third mass is at rest at a distance away. When the simulation starts, the single mass falls into the the binary pair and then the three-body interaction ensues. In all the three-body sets we consider, the initial binary has semi-major axis $a =$ 5 AU, and eccentricity $e =$ 0 (see Figure~\ref{fig:icscheme}. The initial distance of the single from the binary center of mass is 100 AU. Both the single mass and the center of mass of the binary pair are initially at rest. 

Keeping these parameters constant, we vary over the true longitude $\lambda$ of the initial binary pair and inclination of the single mass with respect to binary plane. The true longitude of binary pair is drawn uniformly from $[0,2\pi]$ and the inclination of single $\theta$ is drawn uniformly from $[0,\pi]$. In this manner, for each three-body system, a set of $10^6$ realizations with different initial configurations are generated. All this information has been summarized in Table~\ref{tab:sims_summary}. 

For the three-body systems under consideration here, the total energy $E$ of the system is given by 
\be
E = -\frac{m_a m_b}{2a} - \frac{m_s(m_a+m_b)}{d},\qquad d = 100\text{AU}, a = 5\text{AU}
\ee
and the total angular momentum $L$ is given by
\be
L = m_a m_b \sqrt{\frac{a (1 - e^2)}{m_a + m_b}}
\ee
where \{$m_a$,$m_b$\} are the masses that initially form the binary pair and $m_s$ is the single mass. As we initially assume a circular binary, we have $e = 0$. Note that in our numerical scattering experiments, we assume for the gravitational constant $G = 1$, thus $G$ does not appear in the above expressions for $E$ and $L$.

\begin{table*}
	\centering
	\caption{Summary of the three-body simulations sets. `$d$' refers to the distance between single and binary center of mass. `$N$' refers to the total number of realizations in a given simulation set. $|E|$ and $L$ are, respectively, the absolute value of the total energy and the total angular momentum of the three-body system. Thus, each simulation set has $10^6$ different three-body interactions. The mass contrast $\delta$ (defined in Section~\ref{ssec:delta}) is a quantitative indicator of how unequal the mass distribution is in the three-body system.}
	\label{tab:sims_summary}
	\begin{tabular}{cccccccccc} 
		\hline
		Masses($M_{\odot}$) & Binary(\Mdot) & Single(\Mdot) & Binary semi-major Axis (AU) & Binary Eccentricity & d (AU) & N & $\delta$ & $|E|$ & $L$\\
		\hline
		15,15,15 & 15,15 & 15 & 5 & 0 & 100 & $10^6$ & 1.00 & 27.00  & 91.85 \\
		12.5,15,17.5 & 15,17.5 & 12.5 & 5 & 0 & 100 & $10^6$ & 3.54 & 30.31 & 102.96\\
		12,15,18 & 15,18 & 12 & 5 & 0 & 100 & $10^6$ & 4.59 & 30.96 & 105.09\\
		10,10,20 & 10,20 & 10 & 5 & 0 & 100 & $10^6$ & 12.32 & 23.00 & 81.64\\
		10,15,20 & 15,20 & 10 & 5 & 0 & 100 & $10^6$ & 13.66 & 33.50 & 113.38\\
		10,20,20 & 20,20 & 10 & 5 & 0 & 100 & $10^6$ & 14.70 & 44.00 & 141.42\\
		8,21,21 & 21,21 & 8 & 5 & 0 & 100 & $10^6$ & 44.14 & 47.46  & 152.15\\
		5,15,25 & 15,25 & 5 & 5 & 0 & 100 & $10^6$ & 494.10 & 39.5 & 132.58 \\
		\hline
	\end{tabular}
\end{table*}

\subsection{Numerical errors}

As discussed earlier, the three-body system is an example of a chaotic system where small perturbations in initial conditions will lead to exponentially different final outcomes. Therefore, numerical round-off errors in initial conditions will eventually lead to exponentially different solutions from the \textit{true} solution. In our numerical simulations, the cumulative errors in total energy $E$ and total angular momentum $\vec{L}$ conservation are approximately $10^{-12} \sim 10^{-13}$. However, even with these low errors, the small differences are going to exponentially manifest into differences between integrated three-body evolution and true, converged evolution. Therefore, most N-body codes are not going to produce individually accurate and precise three-body solutions that are time-reversible in chaotic parts of the phase space even when $E$ and $L$ are highly conserved. This accumulation of round-off errors alone is enough to make individual solutions unreliable. 

Using reliable three-body simulations is essential to do a robust comparison between theory and simulations. The unreliability of individual three-body simulations discussed above might prove to be an issue. However, the errors are sufficiently small that the outcome distribution functions seen in the simulations are converged. Numerous studies have in fact shown that even though few- and N-body codes do not produce individually accurate solutions, their results are correct in an ensemble sense. That is, statistically the simulations are true representations of the N-body problem  \citep[e.g.][]{tjarda15,goodman93,smith79}. Therefore, our numerical simulations are reliable for assessing statistical properties of few-body chaos, even if an individual solution in the chaotic sea is not trustworthy.

\subsection{Three-Body Evolution}
\label{ssec:tbp_evols}

As discussed in Section~\ref{simexp}, three-body interactions in our simulation setup start with a binary-single scattering as the initial single mass particle falls into the binary pair from a large distance. This, therefore, does not allow for formation of stable triple systems due to the time-reversible nature of Newton's equations. That is, if a three-body system were stable, then it would have to be stable at all times in future and past, but that is clearly not true in the beginning. Therefore, as discussed in \citet{hut83_topology} as well, a solution to an unstable three-body system involves a scattering event(s) where the system becomes simple both in the future ($t \rightarrow \infty $) and in the past ($t \rightarrow -\infty$). Hence, each three-body system decays into a binary pair and a single mass that escapes to infinity. Note that as we only consider three-body systems with a negative total energy, three-body ionization or a triple breakup is not possible. The 3 different kinds of three-body end states can be easily identified by which of the 3 masses is ejected after the system decays. We also neglect interactions that would result in a collision between particles. They represent a very small fraction of the entire ensemble of interactions ($\sim$5-6 out of $10^6$ simulations) and therefore removing them does not impact the statistics of the ensemble. 

From the initial in-fall of the single mass to the termination of the three-body interaction, the entire duration is defined as the lifetime or decay time of the three-body system. During its entire evolution, the system mainly exists in 2 states: scramble state and sub-escape excursion state. A scramble state is when there is no well-defined hierarchy or binary pair in the system. Therefore, in a scramble state the 3 masses are roughly equidistant from each other and in approximate energy equipartition. As discussed in \citet{stone19} and \citet{manwadkar20}, the scramble state is the key dynamical state that ergodicizes the three-body system. By looking at movies of individual three-body interactions in our simulations, we find that a single scramble lasts $\sim$2-3yrs. The notion of a scramble will be very useful in defining our ergodic cut later in Section~\ref{thecut}. A sub-escape excursion state is when there is a well-defined binary pair and a single that goes on a temporary, wide orbit. During a sub-escape, one of the particles gets enough energy to leave the binary pair temporarily, hence going on a wide orbit. However, as the single particle is still bound to the binary, it eventually returns for a close encounter. At any given point in time in a general three-body interaction, the system exists in one of these 2 states. Depending on which kind of state dominates the three-body evolution, one can find: (i) extremely long-lived interactions dominated by sub-escapes that occupy the power-law tail end of the lifetime distribution, (ii) relatively short-lived interactions dominated by scrambles that occupy the exponential part of the lifetime distribution. In addition to this, one can find very short-lived three body interactions with prompt ejections. In these interactions, immediately after the initial in-fall of the single particle into the binary pair, one of the particles gains enough energy to escape to infinity, immediately terminating the interaction. The lifetime of such interactions is, therefore, just the initial in-fall time. In our simulations, the in-fall time of the single particle is $\sim$27yrs. For more details on the different kinds of three-body interactions and their properties refer to \citet{manwadkar20}.

\section{Quantitative Metrics}
\label{sec:quant_metrics}
\subsection{Mass Contrast}
\label{ssec:delta}

We find it useful to define a measure of mass inequality within a mass set, which we call ``mass contrast'' denoted by $\delta$. It is given by the ratio of maximal and minimal expected escape probabilities, which correspond to the lightest and heaviest objects, respectively, namely
\begin{equation}
\delta = \frac{P\textsubscript{lightest}}{P\textsubscript{heaviest}} = \left( \frac{k_L}{k_H}  \right)^{1.5}
\end{equation}
where $P\textsubscript{lightest}$ and $P\textsubscript{heaviest}$ are the escape probabilities of the lightest and heaviest masses in the three-body system respectively,  
and $k_L,k_H$ are the corresponding binary constants defined in \eqref{def:k}. 
As we know the analytical expression for the escape probability, we can compute the expression for the mass contrast $\delta$ as a function of $m_1,m_2$ and $m_3$. The convention is that $m_1$ is the lightest mass and $m_3$ is the heaviest mass. 
\begin{align}
\delta &= \left(\frac{k_L}{k_H}\right)^{1.5} = \left(\frac{k_1}{k_3}\right)^{1.5} \\
&= \left[\frac{(m_2 m_3)^3}{m_2 + m_3} \frac{m_1 + m_2}{(m_1 m_2)^3}  \right]^{1.5} \\
&= \left[\left(\frac{m_3}{m_1}\right)^{3} \frac{m_2 + m_1}{m_2 + m_3}  \right]^{1.5}
\end{align}

Thus, the expression for the mass contrast is,
\begin{equation}
\delta(m_1,m_2,m_3) = \frac{P\textsubscript{lightest}}{P\textsubscript{heaviest}} = \left[\left(\frac{m_3}{m_1}\right)^{3} \frac{m_2 + m_1}{m_2 + m_3}  \right]^{1.5}
\end{equation}
For equal mass systems, we have $\delta = 1$. While for more extreme mass ratio systems like $\{5,15,25\}$, we have $\delta = 494.1$. The mass contrast for each system under consideration is given in Table~\ref{tab:sims_summary}.

One notices that for a fixed $m_2$, as $\frac{m_3}{m_1} \rightarrow \infty$, we have $\delta \rightarrow \infty$. Therefore, the more unequal the mass distribution, the higher is value of the mass contrast $\delta$. 

\subsection[The Gap Time]{The Gap Time $\tau\textsubscript{gap}$}
\label{ssec:gap_time}
The gap time $\tau\textsubscript{gap}$ is defined as the time interval between the last sub-escape in a three-body interaction and the ejection of one of the particles to infinity terminating the interaction. If $\tau\textsubscript{gap} > 0$, the three-body system existed in a scramble-state before the ejection. Therefore, the ejection of one of the particles happened from a scramble state and not as a prompt ejection after one of the particles returns from the sub-escape. This gap time $\tau\textsubscript{gap}$ will be crucial in identifying the subset of three-body interactions that have an ergodic disintegration, i.e., an ejection from the ergodic regime.

\begin{figure*}
    \centering
        \begin{subfigure}[b]{\textwidth}
            \centering
            \includegraphics[width=\textwidth]{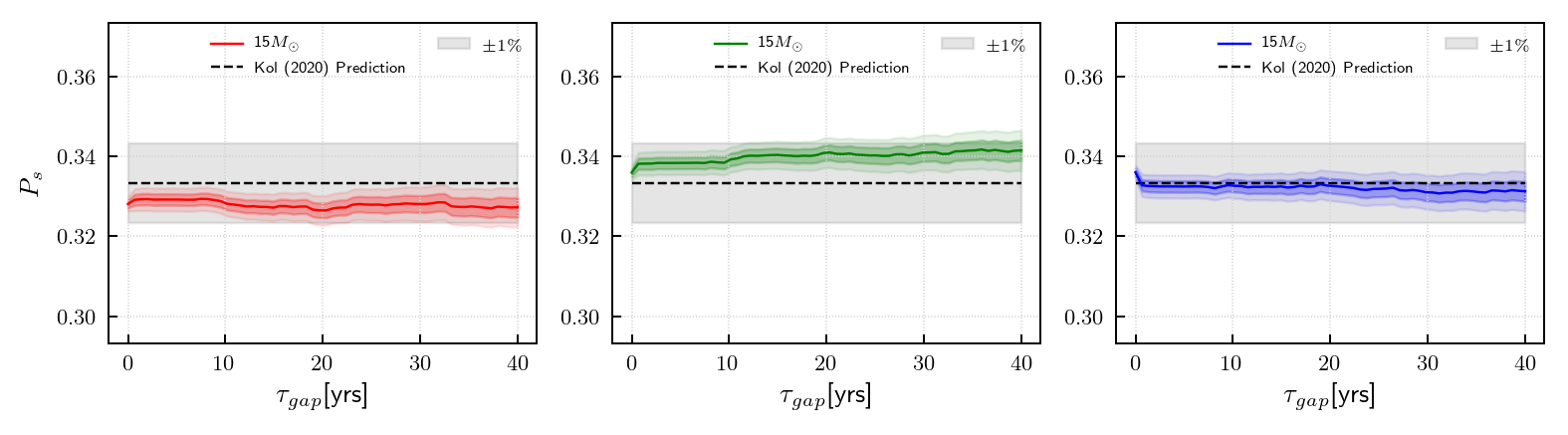}
            \caption[]%
            {{\small 15,15,15\Mdot system}}    
        \end{subfigure}
        \begin{subfigure}[b]{\textwidth}  
            \centering 
            \includegraphics[width=\textwidth]{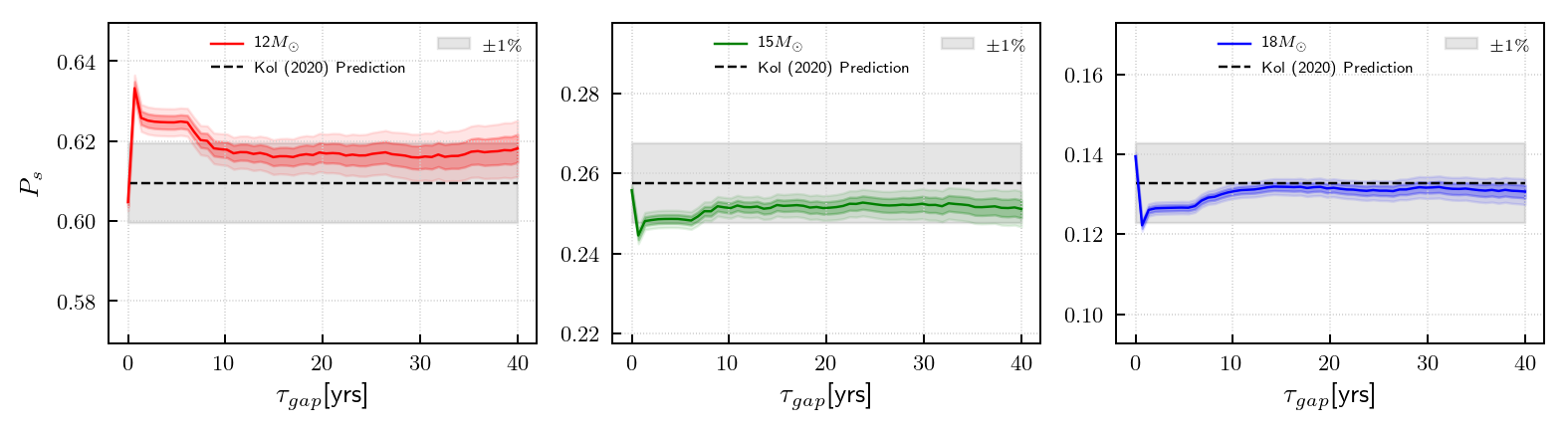}
            \caption[]%
            {{\small 12,15,18\Mdot system}}  
        \end{subfigure}
        \begin{subfigure}[b]{\textwidth}  
            \centering 
            \includegraphics[width=\textwidth]{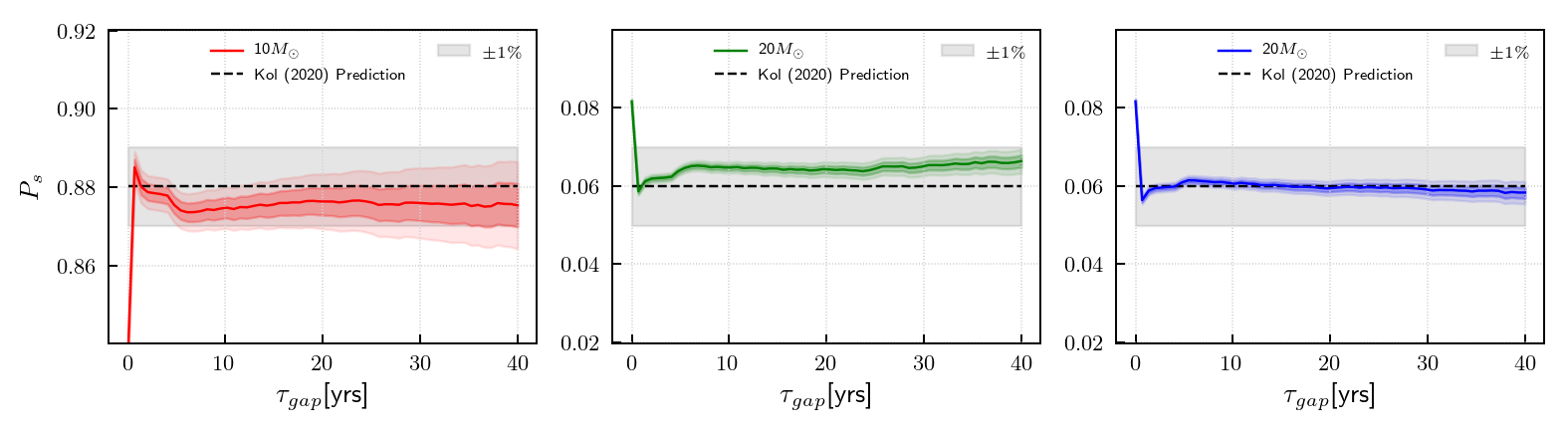}
            \caption[]%
            {{\small 10,20,20\Mdot system}}    
        \end{subfigure}
\caption{The escape probabilities $P_s$ as a function for different ergodic cuts by varying the $\tau\textsubscript{gap}$ cutoff for 3 different three-body systems: (a) 15,15,15 \Mdot (b) 12,15,18\Mdot (c) 10,20,20\Mdot. In all the cases, the lifetime cut of $\tau_D > 80$yrs is applied. The horizontal black dashed line shows the theoretical prediction of $P_s$ value from \citet{Kol2020}. The gray shaded region is the $\pm 1\%$ region around the theoretical prediction for $P_s$. The dark- and light-colored shaded regions around the solid colored line denote the $1\sigma$ and $2\sigma$ uncertainties in ejection probabilities in simulations. The uncertainties are calculated using standard Poisson errors. Similar plots for the other three-body systems under consideration are shown in the Appendix~\ref{appendix:erg_cut}.}
\label{fig:tau_con1}
\end{figure*}

\begin{table*}
\caption{The escape probabilities $P_s$ and total number of realizations $N_s$ by escaper mass in different three-body systems. The total number of simulations in each set is $10^6$. The cut-offs are as follows: [1] The entire set, [2] Removing prompt ejections ($\tau\textsubscript{D} > 30$yrs),
[3] ergodic subset ($\tau\textsubscript{gap} > 10$yrs and $\tau\textsubscript{D} > 80$yrs). The ergodic subset (Cut 3) corresponds to the interactions where the final escape happens from the ergodic regime. For discussion on this cut, refer to Section~\ref{thecut}. The probabilities of the ergodic subset are being compared to theoretical formalisms: \citet{Kol2020}: K20, \citet{ginat20} : GP20, \citet{stone19}: SL19, and \citet{Valtonen_book_2006}: VK06. For the equal mass case, the initial binary pair masses are denoted by (1) and (2).}
\label{table:eject_prob}
\centering
\begin{tabular}{ccccccaaccc}
\hline
Masses($M_{\odot}$) & Ejection Mass($M_{\odot}$) & $P_{s}$ [1] & $N_s$ [2] & $P_{s}$ [2] & $N_s$ [3] & $P_{s}$ [3] & K20 & GP20 & SL19  & VK06 \\
\hline
& 15 (0) & 0.199  &199045 & 0.329 & 45208 &  0.328 &0.333 & 0.333 & 0.333 & 0.333 \\
15,15,15 & 15 (1) & 0.400 &203155& 0.336 & 46655 & 0.339 & 0.333 & 0.333 & 0.333 & 0.333\\
 & 15 (2) & 0.401 &203159& 0.335 & 45771 & 0.333 &0.333 & 0.333 & 0.333 & 0.333 \\
\hline
 & 12.5  & 0.433  & 433047 & 0.560 & 89428 & 0.572 & 0.562 & 0.504 & 0.639 & 0.492  \\
12.5,15,17.5 & 15& 0.302 & 201393 & 0.261 & 42525 & 0.272 & 0.279 &0.297&0.247& 0.305  \\
 & 17.5  & 0.265 & 138310 & 0.179 & 24429 & 0.156 & 0.159 & 0.199 & 0.114 & 0.203 \\
\hline
 & 12  & 0.503  & 503471 & 0.608 & 98807  & 0.618 & 0.610 & 0.542 & - & 0.525  \\
12,15,18 & 15 & 0.283 & 198769 & 0.240 & 40136 & 0.251 & 0.257 & 0.281 & - & 0.293  \\
 & 18  & 0.214 & 126009 & 0.152 & 20881 & 0.131 &  0.133 & 0.177 & - & 0.182 \\
\hline
 & 10  & 0.530  & 524093 &  0.554 & 67278  &  0.4736 & 0.4805 & 0.4604 & - & 0.463 \\
10,10,20 & 10 &  0.418 & 37013 & 0.392  & 69555 & 0.4896 &  0.4805 & 0.4604  & - & 0.463  \\
 & 20 &  0.052 & 50791 & 0.054 & 5233 & 0.0368 & 0.0390 & 0.0792 & - & 0.074 \\
\hline
 & 10  & 0.727  &726841& 0.734 & 113627 & 0.784 & 0.784 & 0.697 & 0.872 & 0.649  \\
10,15,20 & 15  & 0.193 &182416& 0.184 & 22988 & 0.159 & 0.159 & 0.202 & 0.103 &  0.236 \\
 & 20 & 0.080  &80496& 0.081 & 8306 &0.057 & 0.057 & 0.101 & 0.025 & 0.115\\
\hline
 & 10  & 0.842  & 803550 & 0.838  &  108978 & 0.875 &0.880 & 0.796 & - & 0.726 \\
10,20,20 & 20 &  0.079 & 78019 & 0.081 & 8055 &0.065  &0.060 & 0.102 & - & 0.137  \\
& 20 &  0.079 & 77933 & 0.081 & 7577 & 0.060 & 0.060 & 0.102 & - & 0.137 \\
 \hline
 & 8  & 0.922  & 862162 &  0.920 & 86959  & 0.943 & 0.956 & 0.904 & - & 0.8158 \\
8,21,21 & 21 &  0.039 & 37432 & 0.040 & 2813 & 0.031 & 0.022 & 0.048 & - & 0.0921  \\
 & 21 &  0.039 & 37675 &  0.040 & 2448 & 0.026 & 0.022 & 0.048 & - & 0.0921 \\
 \hline
 & 5  & 0.9712  & 826963 &0.9664  & 41960  &0.9712 &0.9872 & 0.9667 & - & 0.8796 \\
5,15,25 & 15 &  0.0248 & 24724 &0.0289 & 1108 &0.0256 & 0.0108 & 0.0245 & - & 0.0895  \\
 & 25 &  0.0040 & 4036 & 0.0047 & 137 &0.0032 &0.0020 & 0.0088 & - & 0.0309 \\
\hline
\end{tabular}
\end{table*}

\section{The Ergodic Cut}
\label{thecut}

The theory presented in \citet{Kol2020} gives predictions for outcome statistics from the ergodic regime. To successfully identify the subset of interactions that have `\textit{escape from ergodic regime}', we need to first remove the very short lived interactions (prompt ejections) that demonstrate regular or non-chaotic behavior. Based on discussion in Section~\ref{ssec:tbp_evols}, a cut of $\tau_D > 30$yrs will remove the prompt ejections. However, interactions that are longer than $30$yrs can still demonstrate regular or non-chaotic behavior. For instance, looking at initial condition phase space maps of three-body systems in \cite{manwadkar20}, one notices that interactions with $\tau_D < 80$yrs conglomerate in band-like structures in phase space. All the interactions in these bands have the same lifetime, same ejected particle etc. Therefore, tiny perturbations in initial conditions in these band-like structures do not result in an exponentially divergent interaction. Therefore, we first apply a fairly conservative cut of $\tau_D > 80$yrs to consider interactions that are ergodic to some degree. 

In addition to this, we need to apply a cut in $\tau\textsubscript{gap}$ (defined in Section~\ref{ssec:gap_time}) to ensure that the escape happens from a scramble-state that has lasted long enough to have been ergodicized. We observe that after $\sim 3-4$ consecutive scrambles or three-body close encounters, the system has forgotten information about its previous state. As mentioned in Section~\ref{ssec:tbp_evols}, each scramble state lasts for approximately $\sim2-3$yrs. Therefore, we apply a cut of $\tau\textsubscript{gap} > 10$yrs to filter out the interactions that have an escape from the ergodic regime. To demonstrate convergence in our criterion, Figure~\ref{fig:tau_con1} shows the escape probabilities of the remaining subset after applying the $\tau_D > 80$yrs cut and the cut in $\tau\textsubscript{gap}$. We notice that after $\tau\textsubscript{gap} = 10$yrs, the escape probabilities do not change appreciably suggesting that our cut is a converged cut and applying more strict cuts does not influence the statistics. In Appendix~\ref{appendix:erg_cut}, we show similar plots for other three-body systems where we observe similar converged behavior. However, in the 8,21,21 system, we do not observe converged behavior. This is likely due to very low number statistics. 

Therefore, to obtain the subset of three-body interactions that have `escape from ergodic regime', we apply the following cuts: (i) $\tau_D > 80$yrs and (ii) $\tau\textsubscript{gap} > 10$yrs. 

Note that there is an important distinction to be made between an `\textit{ergodic interaction}' and `\textit{escape from ergodic regime}'. The former suggests that the entire interaction is ergodic as it is dominated by scramble-states and thus its entire evolution occurred in the ergodic region of phase space. The latter suggests that regardless of the state of the three-body interaction sufficiently prior to ejection, the final ejection happened from a scramble state that has been sufficiently ergodicized. This criterion for ``sufficiently ergodicized'' has been established earlier as at least 3 scrambles which corresponds to $\tau\textsubscript{gap} > 10$yrs. Therefore, an `\textit{escape from ergodic regime}' does not imply `\textit{ergodic interaction}'. For comparing simulation outcome statistics to \citet{Kol2020} theoretical predictions for ergodic escape, we will focus on the interactions that have '\textit{escape from ergodic regime}'.

\section{Comparisons to Numerical Scattering Experiments}
\label{results}

In this section we present our comparisons between the theoretical predictions of \citet{Kol2020} and the three-body numerical simulations.

\subsection{Ergodic Escape Probability}

{ 
\renewcommand{\arraystretch}{1.6}
\begin{table*}
\caption{The fitted parameters $\alpha$ and $l_{F,c}$ (denoted by [1] in this table) fitted to the near-threshold distribution of effective system angular momenta $l_{F}$ by escaper mass for different three-body systems.
The uncertainties shown are the 2$\sigma$ error bars. We should be expecting $\alpha$ in \citet{Kol2020} is $\alpha = 2$ and critical value $l_{F,c}$ according to Equation~\ref{eqn:Lfc}. The near-threshold distribution is determined by interactions with the smallest x\% $l_F$ values where $x$ is given in the Data Range column. $N$ column shows the number of interactions that are considered as part of the near-threshold distribution. The ``$l_{F,c}$ [K20]'' column contains the predicted values of $l_{F,c}$ from \citet{Kol2020} and the ``$l_{F,c}$ [Simulations]'' column contains the minimum value of $l_{F}$ found in simulations. }
\label{table:lf_table}
\centering

\begin{tabular}{ccccccccc}
\hline
Masses($M_{\odot}$) & Ejection Mass($M_{\odot}$)  & Data Range & N & $\alpha$ [1] & $l_{F,c}$ [1] & $l_{F,c}$ [K20] & $l_{F,c}$ [Simulations]  \\
\hline
& 15 (0)  & 2\% & 905  & $1.72^{+0.12}_{-0.12}$ & $9.53^{+0.67}_{-0.69}$ &  8.00 & 13.01 \\
15,15,15 & 15 (1) &2\%&934  & $1.97^{+0.14}_{-0.13}$ & $9.68^{+0.71}_{-0.74}$ & 8.00 & 11.99 \\
 & 15 (2) &2\%&916 & $2.06^{+0.15}_{-0.14}$ & $10.07^{+0.76}_{-0.78}$ & 8.00 & 11.56 \\
\hline
 & 12.5 &2\%&1789  & $2.18^{+0.10}_{-0.10}$ & $6.79^{+0.52}_{-0.53}$ & 7.14  & 8.92  \\
12.5,15,17.5 & 15 &2\%&851  & $1.85^{+0.35}_{-0.29}$ & $27.65^{+1.16}_{-1.57}$ & 27.09 & 28.97  \\
 & 17.5  &2\%&489 & $2.27^{+0.57}_{-0.51}$ & $40.14^{+2.03}_{-2.21}$ & 40.08 & 42.41   \\
\hline
 & 12  &2\%&1977   & $1.92^{+0.09}_{-0.09}$ & $7.29^{+0.45}_{-0.47}$ & 6.95  & 9.32  \\
12,15,18 & 15 &2\%&803  & $2.01^{+0.37}_{-0.20}$ & $32.21^{+0.50}_{-1.50}$ &  31.44 & 32.74   \\
 & 18  &2\%&418 & $1.93^{+0.81}_{-0.47}$ & $47.66^{+1.31}_{-2.83}$ & 46.03 & 49.03   \\
\hline
 & 10  &2\%&1346 & $1.60^{+0.10}_{-0.10}$ & $6.00^{+0.33}_{-0.34}$ & 5.51  & 7.12   \\
10,10,20 & 10 &2\%& 1392 & $1.56^{+0.10}_{-0.10}$ & $6.59^{+0.31}_{-0.35}$  & 5.51 & 6.99   \\
 & 20 & 5\% &262  & $2.04^{+0.85}_{-0.66}$ & $49.36^{+1.07}_{-2.39}$ &  48.68 & 50.49   \\
\hline
 & 10  &2\%&2273 & $1.93^{+0.09}_{-0.08}$ & $6.38^{+0.36}_{-0.38}$ & 6.08  & 7.50   \\
10,15,20 & 15  &2\%& 460  & $2.32^{+0.55}_{-0.53}$ & $49.36^{+1.53}_{-1.46}$ & 50.30 & 50.99 \\
 & 20 & 5\% &416 & $1.95^{+0.76}_{-0.65}$ & $68.61^{+2.31}_{-2.42}$ & 68.50  & 71.11  \\
\hline
 & 10  & 2\%& 2180  & $1.93^{+0.09}_{-0.09}$ & $7.24^{+0.42}_{-0.43}$  &6.58 & 8.16  \\
10,20,20 & 20 & 5\%& 403 & $2.42^{+0.46}_{-0.65}$ & $86.19^{+2.22}_{-2.01}$ & 86.37 & 88.59  \\
& 20 &5\% & 378 & $2.18^{+0.65}_{-0.73}$ & $86.97^{+2.56}_{-2.69}$ & 86.37 & 89.71  \\
 \hline
 & 8  & 2\%& 870& $2.02^{+0.13}_{-0.13}$ & $5.31^{+0.41}_{-0.42}$ & 5.48 & 5.99  \\
8,21,21 & 21 & 5\% & 282 & $2.32^{+0.62}_{-0.90}$ & $111.64^{+2.35}_{-2.94}$ & 110.65 & 114.12  \\
 & 21 & 5\%& 245 & $2.16^{+0.69}_{-0.83}$ & $110.53^{+2.24}_{-2.01}$ & 110.65 & 112.92 \\
\hline
 & 5  & 2\%& 840   & $1.78^{+0.13}_{-0.12}$ & $4.22^{+0.30}_{-0.31}$ & 3.40 & 4.95  \\
5,15,25 & 15 & 10\% & 111  & $2.43^{+0.54}_{-1.08}$ & $103.49^{+1.44}_{-1.97}$ & 103.87  & 105.01  \\
 & 25 & 20\% & 28  & $2.17^{+0.78}_{-1.08}$ & $117.69^{+2.29}_{-3.52}$ & 116.24 & 120.09  \\
\hline
\end{tabular}
\end{table*}
} 

As each three-body system in our setup disintegrates into a final binary pair and a single mass, there are 3 different ways in which a three-body system can disintegrate. We calculate the escape probability $P_s$ in our simulations as
\be
P_s = \frac{N_s}{N}
\ee
where $N$ is the total number of simulations that satisfy the ergodic escape cut and $N_s$ is the total number of simulations in that set that result in the particle $m_s$ being ejected. As discussed earlier in Section~\ref{pert_pred}, the theoretical prediction for the escape probability $P_s$ in the \citet{Kol2020} ergodic formalism is given by
\be
P_s \propto k_s^{3/2}
\ee
In Table~\ref{table:eject_prob}, the two grey columns show the relevant comparisons between the simulation escape probabilities and the \citet{Kol2020} theoretical predictions. For an additional comparison, we also present the predicted escape probabilities from three other three-body ergodic formalisms presented in \citet{ginat20}, \citet{stone19} and \citet{Valtonen_book_2006}. The escape probability according to \citet{Valtonen_book_2006} is 
\be 
P^{\text{VK06}}_s \propto m^{-q}_{s}
\ee
where $q$ is a function of total angular momentum $L$. In our three-body experiments, we find $q$ to be in the range $2.1 - 2.7$. Refer to Appendix~\ref{appendix:VK06_ps} for a detailed description of this dependence and calculation. The escape probability according to \citet{ginat20} is given by Equation 32 in their paper which is 
\be 
P^{\text{GP20}}_s \propto \frac{m^{4}_a m^4_b}{(m_a+m_b)^{5/2}}
\ee
For \citet{stone19}, the probabilities are obtained by integrating over the outcome distribution, which is given there in closed form.  

Table~\ref{table:eject_prob} shows the escape probabilities corresponding to three different subsets of numerical simulations and the comparison with the above discussed theoretical predictions. $P_s[1]$ are the escape probabilities for the entire simulation set. These will be affected by the initial setup of the simulations as this set contains prompt ejections. For instance, in the $15,15,15$\Mdot system, the escape probabilities for entire set are $0.199,0.400,0.401$ respectively. One can infer that the first mass is the single mass and the latter two are the initial binary pair. $P_s[2]$ are the escape probabilities for interactions that are not prompt ejections, that is, $\tau_D > 30$yrs. $P_s[3]$ are the escape probabilities for the interactions that have `escape from ergodic regime'. This subset is obtained by applying the ergodic cut detailed in Section~\ref{thecut}. It is these ergodic escape probabilities, i.e. $P_s[3]$, that are supposed to be compared with the theoretical formalisms. It is fascinating to see that the theoretical predictions from \citet{Kol2020} agree down to the 1\% level with $P_s[3]$. No other theoretical formalism is that accurate in predicting the escape probabilities. 

In conclusion, Table~\ref{table:eject_prob} shows the excellent agreement between the simulation ergodic escape probabilities (i.e. $P_s[3]$) and the \citet{Kol2020} predictions. Note the big leap in accuracy for the theoretical predictions for \citet{Kol2020} compared to the previous 3 formalisms.

\subsection[Effective Angular Momentum threshold regime]{Low-$l_F$ regime}

\begin{figure*}
	\includegraphics[width=\textwidth]{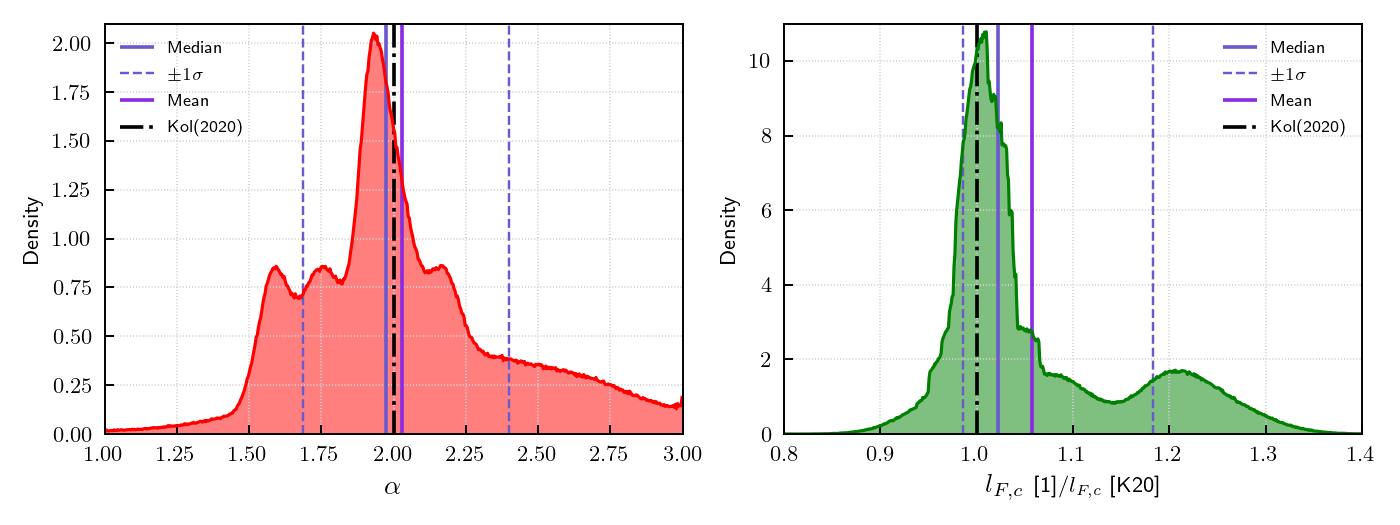}
    \caption{The distribution of the $\alpha$ and $l_{F,c}$ parameter (normalized to \citet{Kol2020} prediction) obtained from the MCMC fitting of $l_F$ threshold distribution for all the three-body systems under consideration. The $l_{F,c}$[1] denotes the value of $l_{F,c}$ obtained through the MCMC fitting (refer to Table~\ref{table:lf_table}. The median value with $1\sigma$ errors for these 2 parameters are $\alpha = 1.97^{+0.42}_{-0.29}$ and $l_{F,c}\text{[1]}/l_{F,c}\text{[K20]} = 1.02^{+0.16}_{-0.04}$. }
    \label{fig:lfc_alpha_dist}
\end{figure*}

As discussed earlier in Section~\ref{theory}, $l_F$ is the angular momentum of the free out-going motion. It can be calculated as follows
\be
l_F = \mu \sqrt{G M (1 - e_h^2) a_h^2}
\ee
where $M = m_B + m_s$, $\mu = \frac{m_B m_s}{M}$ and $e_h$ and $a_h$ are the eccentricity and semi-major axis of the hyperbolic trajectory of the escaping mass. $m_B$ and $m_s$ are the masses of the final binary and the ejected single particle, respectively. The threshold or minimum value of $l_F$ possible is denoted by $l_{F,c}$ and is given by
\be
l_{F,c} := L - \sqrt{\frac{k_s}{(-2E)}  }
\label{eqn:Lfc}
\ee
The values for $l_{F,c}$ for each system and the corresponding ejection type are shown in Table~\ref{table:lf_table}.

\citet{Kol2020} predicts that near the $l_F$ threshold, the $l_F$ probability distribution is independent of absorptivity $\calE$. In this limit, the predicted distribution is of the form
\be
dP_s \propto  \left( l_F- l_{F,c} \right)_+^2 \, dl_F
\ee
We can write the above distribution in its general form as 
\be
\label{eqn:lf_pred}
dP_s = \beta \left( l_F- l_{F,c} \right)_+^\alpha \, dl_F
\ee
where the prediction is that $\alpha = 2$ and the critical value is $l_{F,c}$, defined in Equation~\ref{eqn:Lfc}. Now, to test this prediction, we will look at the distribution of $l_F$ values in the threshold-limit and see whether it agrees with Equation~\ref{eqn:lf_pred}. As we have a finite number of instances, we will have to bin the data to construct the differential distribution. To avoid the issue of discretization due to binning, we will write Equation~\ref{eqn:lf_pred} in a series format\footnote{Appendix~\ref{appendix:lf_series} contains the full derivation of the series formula.} as
\be
\label{eqn:lf_series}
l_{F,n} = l_{F,c} + \gamma n ^{\frac{1}{a+1}} \qquad, \gamma =\left(\frac{1+\alpha}{\beta  N_T}\right)^{1/(1+\alpha)}
\ee
where $N_T$ is the total number of $l_F$ values in the series.

To see whether the simulations agree with the theoretical predictions we fit Equation~\ref{eqn:lf_series} using the Markov Chain Monte Carlo (MCMC) method to the series of $l_F$ values in the threshold regime to find the best fitting values for $\alpha, \gamma$ and $l_{F,c}$. The threshold regime is determined by considering the smallest $2\%$ $l_F$ values, however in cases with low statistics, we relax that condition as seen in Table~\ref{table:lf_table}.

The results of the fittings and the corresponding 2$\sigma$ uncertainties for each ejection type for each three-body system are shown in Table~\ref{table:lf_table}. Table~\ref{table:lf_table} shows excellent agreement between theoretical prediction for $l_{F,c}$ and $\alpha$ and the corresponding fitted value. This can be seen more explicitly in Figure~\ref{fig:lfc_alpha_dist}. The left panel in Figure~\ref{fig:lfc_alpha_dist} shows the total posterior distribution of the fitted $\alpha$ parameter. For a comprehensive statistical description of the posterior distributions, the median, the 15.9th and 84.1th percentiles (as $1 \sigma$ errors) and the mean values are depicted. The median value with $1\sigma$ errors is $\alpha = 1.97^{+0.42}_{-0.29}$ and the mean value is $\alpha = 2.03$. This is in excellent agreement with the theoretical prediction of $\alpha = 2$. The right panel in Figure~\ref{fig:lfc_alpha_dist} shows the total posterior distribution of the fitted $l_{F,c}$ parameter normalized to its theoretical prediction. The median value with $1\sigma$ errors is $1.02^{0.16}_{-0.04}$ and the mean value is $1.06$. This is in excellent agreement with the theoretical prediction of 1. 

Therefore, we conclude that there is excellent agreement between our numerical simulations and the \citet{Kol2020} theoretical predictions about the threshold distribution of $l_F$ values. 

The final results of the low-$l_F$ regime comparison have been summarized in Table~\ref{tab:summary_lfc}.

{ 
\renewcommand{\arraystretch}{1.6}
\begin{table}
	\centering
	\caption{Summary of $l_F$ threshold-regime distribution parameters. The error bars for Median value are 1$\sigma$ error bars.}
	\label{tab:summary_lfc}
	\begin{tabular}{cccc}
		\hline
		$l_F$ Parameters & Median & Mean &  \citet{Kol2020} \\
		\hline
		$\alpha$ & $1.97^{+0.42}_{-0.29}$ & 2.03 & 2.00\\
        $l_{F,c}\text{[1]}/l_{F,c}\text{[K20]}$ & $1.02^{+0.16}_{-0.04}$ &1.06 & 1.00 \\
		\hline
	\end{tabular}
\end{table}
}

\begin{figure*}
    \centering
        \begin{subfigure}[b]{0.485\textwidth}  
            \centering 
            \includegraphics[width=\textwidth]{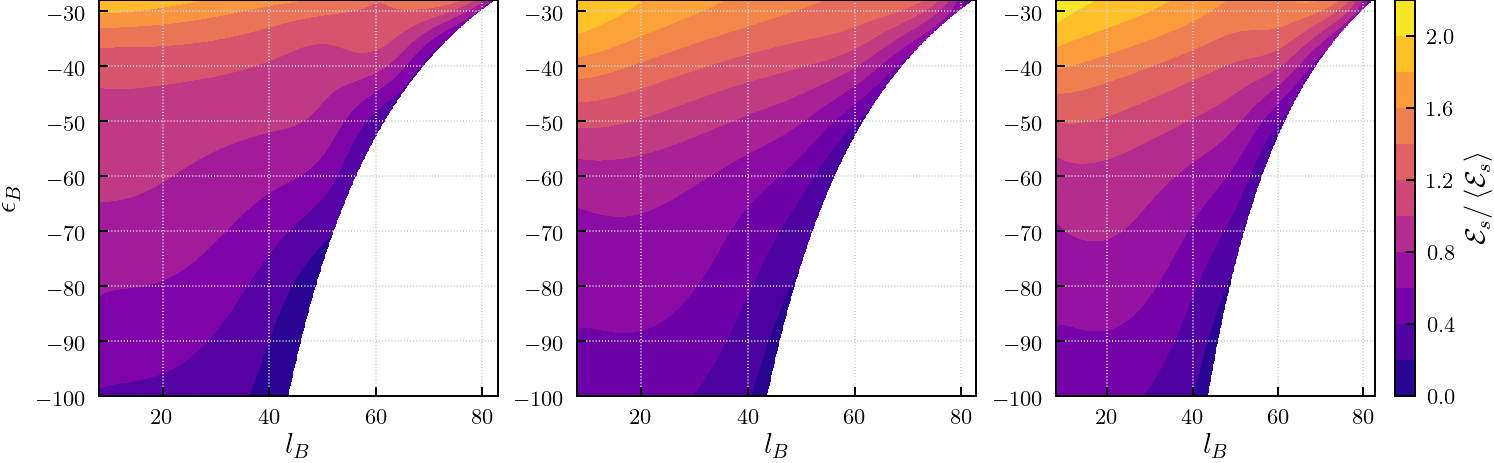}
            \caption[]%
            {{\small  15,15,15 \Mdot system.  }}    
        \end{subfigure}
        \hfill
        \begin{subfigure}[b]{0.485\textwidth}  
            \centering 
            \includegraphics[width=\textwidth]{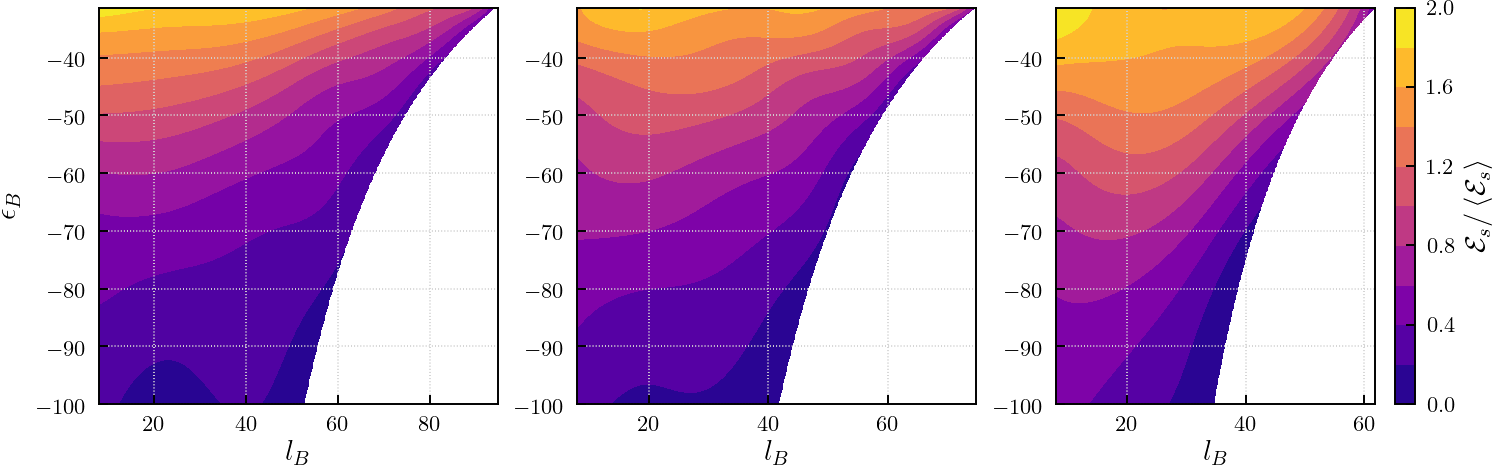}
            \caption[]%
            {{\small 12.5,15,17.5 \Mdot system.  }}    
        \end{subfigure}
        \vskip\baselineskip
        \begin{subfigure}[b]{0.485\textwidth}  
            \centering 
            \includegraphics[width=\textwidth]{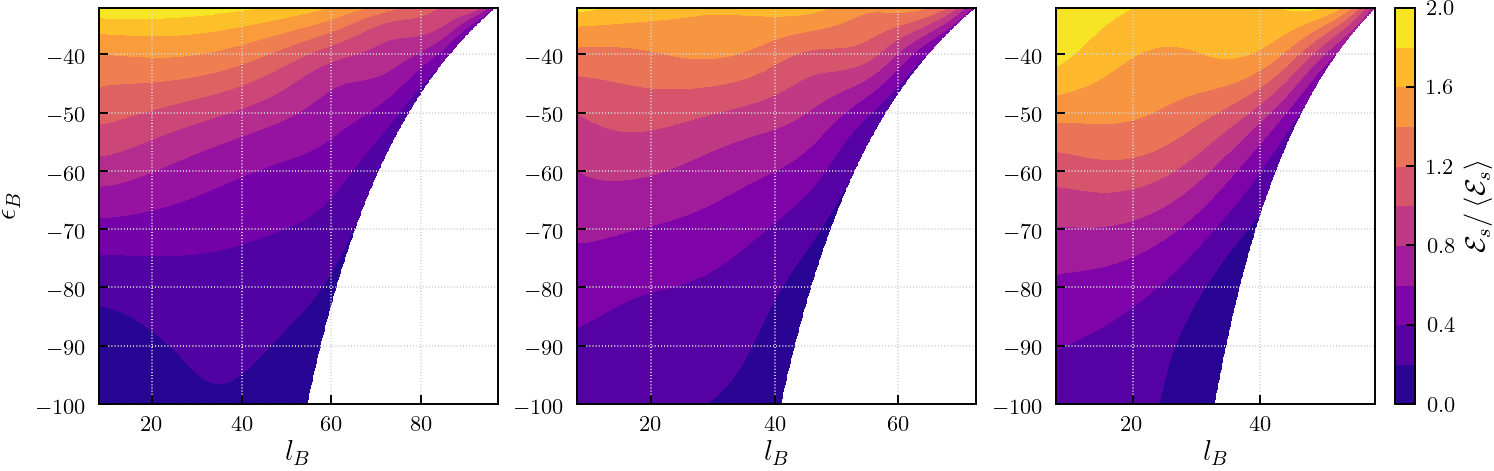}
            \caption[]%
            {{\small 12,15,18 \Mdot system.  }}    
        \end{subfigure}
        \hfill
        \begin{subfigure}[b]{0.485\textwidth}  
            \centering 
            \includegraphics[width=\textwidth]{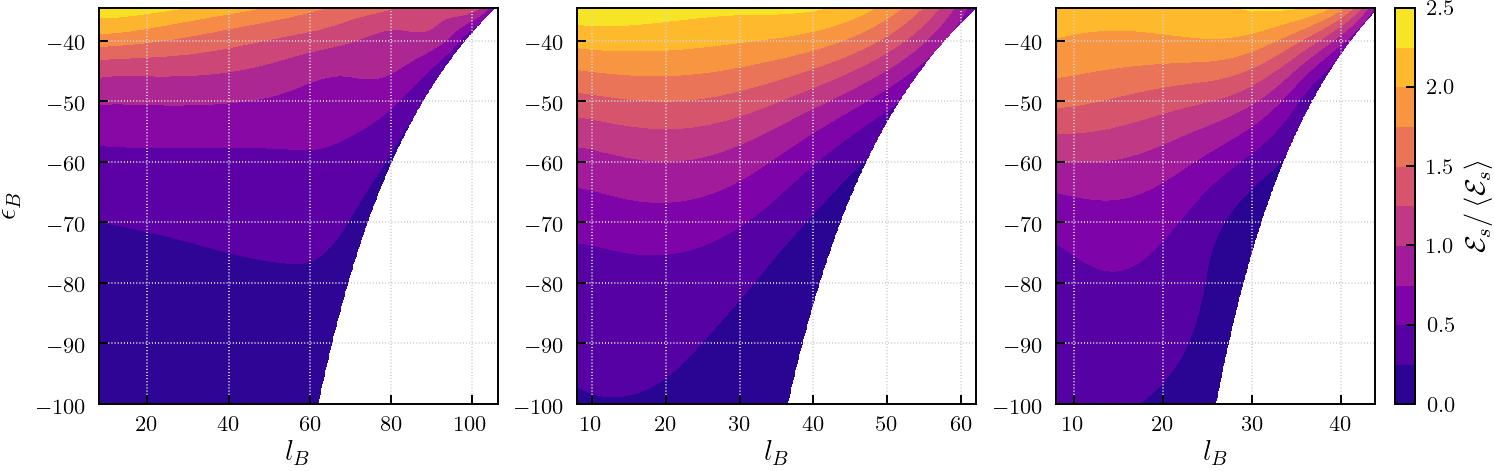}
            \caption[]%
            {{\small 10,15,20 \Mdot system.  }}    
        \end{subfigure}

    \caption{The $\bar{\calE}_{s,\rm pred} / \left<  \calE_{s} \right>$ contour plots for different three-body systems. For each set of 3 contour plots corresponding to a single three-body system, the plots are arranged in ascending order of ejection mass from left to right.}
    \label{fig:cale_nolog}
\end{figure*}

\begin{figure*}
    \centering
        \begin{subfigure}[b]{0.485\textwidth}  
            \centering 
            \includegraphics[width=\textwidth]{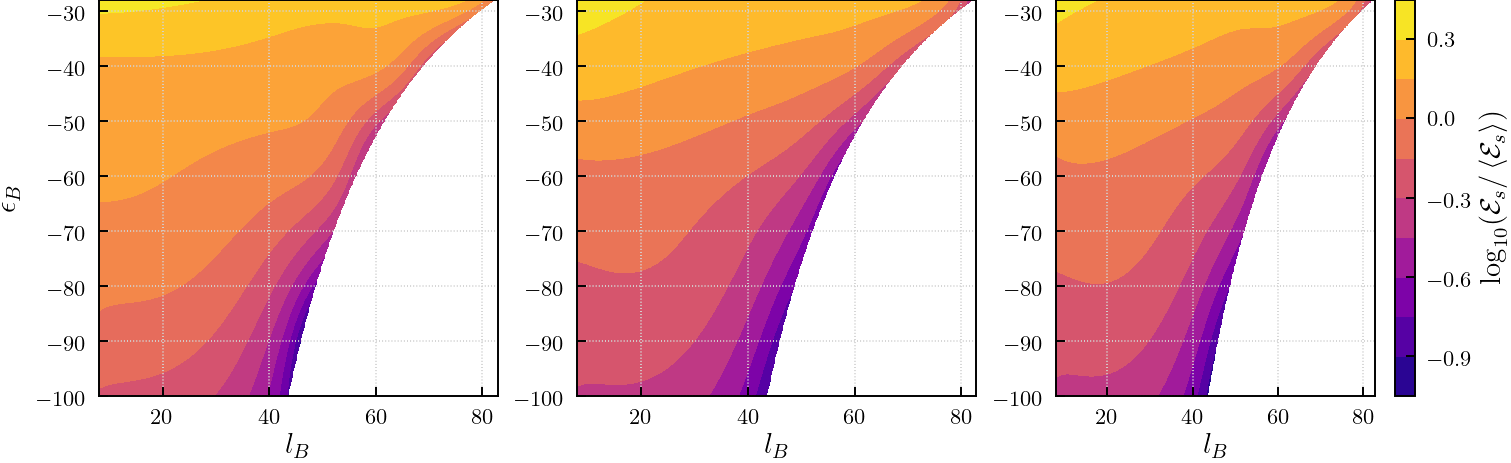}
            \caption[]%
            {{\small  15,15,15 \Mdot system.  }}    
  
        \end{subfigure}
        \hfill
        \begin{subfigure}[b]{0.485\textwidth}  
            \centering 
            \includegraphics[width=\textwidth]{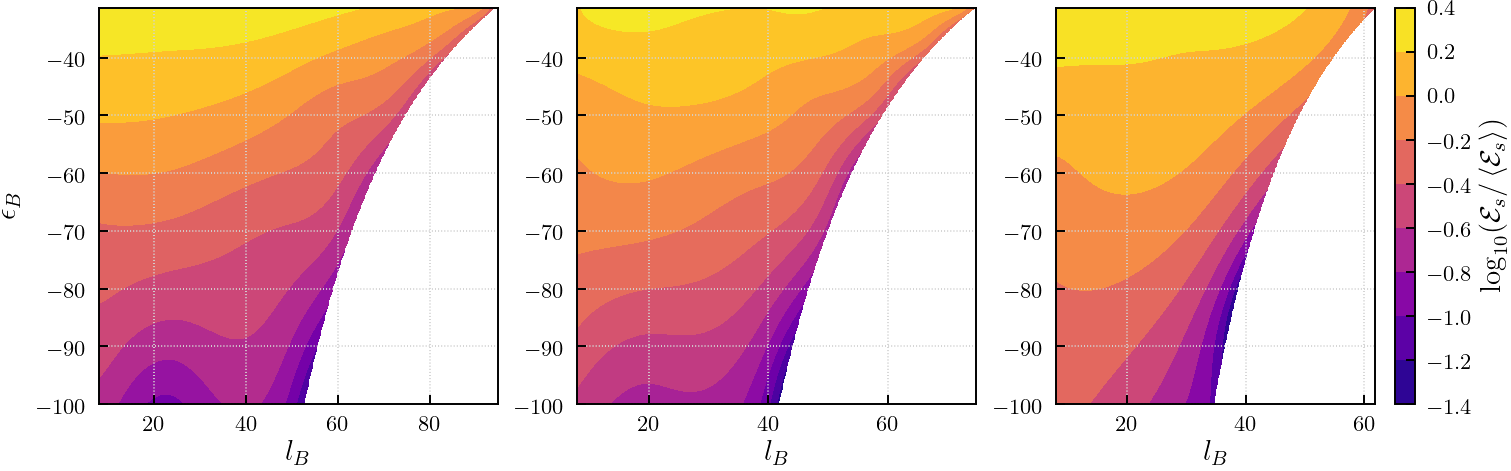}
            \caption[]%
            {{\small 12.5,15,17.5 \Mdot system.  }}    
        
        \end{subfigure}
        \vskip\baselineskip
        \begin{subfigure}[b]{0.485\textwidth}  
            \centering 
            \includegraphics[width=\textwidth]{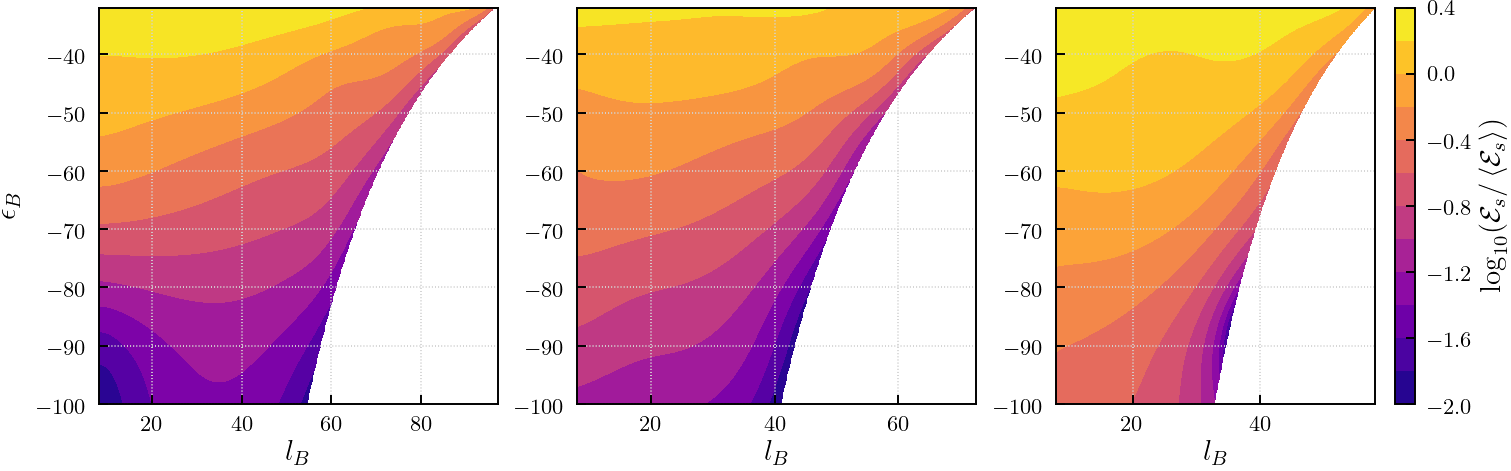}
            \caption[]%
            {{\small 12,15,18 \Mdot system.  }}    
       
        \end{subfigure}
        \hfill
        \begin{subfigure}[b]{0.485\textwidth}  
            \centering 
            \includegraphics[width=\textwidth]{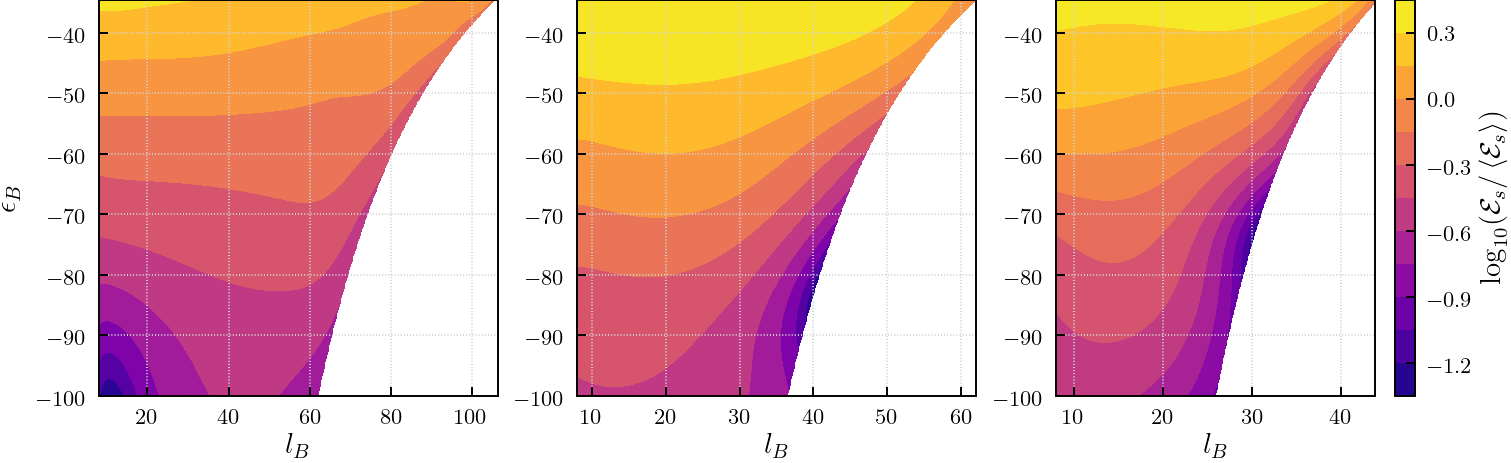}
            \caption[]%
            {{\small 10,15,20 \Mdot system.  }}    
        \end{subfigure}

    \caption{ The $\bar{\calE}_{s,\rm pred} / \left<  \calE_{s} \right>$ log-scaled contour plots for different three-body systems. For each set of 3 contour plots corresponding to a single three-body system, the plots are arranged in ascending order of ejection mass from left to right.}
    \label{fig:cale_log}
\end{figure*}

\subsection{Absorptivity}
\label{ssec: comp_calE}

As discussed earlier, the predicted absorptivity $\bar{\calE}_{\rm s,pred}$ can be measured as 
\be
\bar{\calE}_{\rm s,pred}\left( \eps_B, l_B \right) / \left< \calE _s \right> = \frac{k_s}{6} \frac{(\eps_B/E)^{3/2}}{l_B} \frac{dP_s}{d\eps_B \, dl_B}
 \label{eqn:calE_pred_2}
\ee
where $\left< \calE _s \right>$ denotes a global average of absorptivity, $dP_s=dP_s(\eps_B, l_B)$ is the bi-variate outcome distribution, $E$ is the total energy of system and $k_s$ is the binary constant defined in Equation~\ref{def:k}. 

To compute the bi-variate outcome distribution of binary energy $\eps_B$ and angular momentum $l_B$ from the three-body simulations, we first construct a 2D histogram of $\eps_B,l_B$ for the ergodic subset of interactions. The procedure to obtain the ergodic subset is detailed in Section~\ref{thecut}. By using a bi-variate spline approximation over a rectangular mesh, we smooth the 2D histogram. The 2D histograms and corresponding smoothed bi-variate distributions are shown in Appendix~\ref{appendix:bi_dist}. Using the smoothed bi-variate distributions and Equation~\ref{eqn:calE_pred_2}, we construct the contour plots for absorptivity $\calE$ as seen in Figure~\ref{fig:cale_nolog} and Figure~\ref{fig:cale_log}. Note that the absorptivity contour plots have only been constructed for (i) 15,15,15\Mdot (ii) 12.5,15,17.5\Mdot (iii) 12,15,18 \Mdot and (iv) 10,15,20 \Mdot systems as they have sufficient statistics. 

Looking at Figure~\ref{fig:cale_nolog} and Figure~\ref{fig:cale_log}, we notice that the range of values for $\bar{\calE}_{\rm s} / \left< \calE _s \right>$ is between a minimum of 0 and a maximum of $\approx 2-2.5$, that is, of order 1. This is in agreement with the theoretical prediction presented in Section~\ref{pert_pred}. 

In addition to this, we infer that the absorptivity has a strong dependence on binary energy $\eps_B$ and a weaker dependence on angular momentum $l_B$. Specifically, we find that: (i) for a fixed $l_B$, as we increase the $\eps_B$ (becomes less negative), the absorptivity increases, (ii) for a fixed $\eps_B$, as we $l_B$, the absorptivity decreases slightly. These trends in absorptivity are in agreement with our theoretical understanding of binary-single scattering. As discussed earlier in Section~\ref{theory}, absorptivity $\calE$ is the probability that a scattering of a single off a binary will evolve into a chaotic trajectory rather than a regular scattering like a fly-by or prompt exchange. Therefore, what the absorptivity contour plots suggest is that binaries that have higher energies (soft binaries) have a higher probability to be disrupted into a chaotic interactions. While, binaries that have lower energies (hard binaries) prefer flybys or exchanges. This is in agreement with binary-single scattering understanding in \citet{heggie75} that ``hard binaries harden, and soft binaries soften''. Soft binaries that undergo binary-single scattering will become more soft, resulting in the system being closer to energy equipartition than in the case where the hard binary is hardened. This implies that hard binaries are resilient to being chaotically disrupted by encounters with a single mass, while soft binaries are more likely to be disrupted into a chaotic scramble, thus having higher absorptivity.

\subsection{Lifetime Distribution}

\begin{figure*}
    \centering
        \begin{subfigure}[b]{0.485\textwidth}  
            \centering 
            \includegraphics[width=\textwidth]{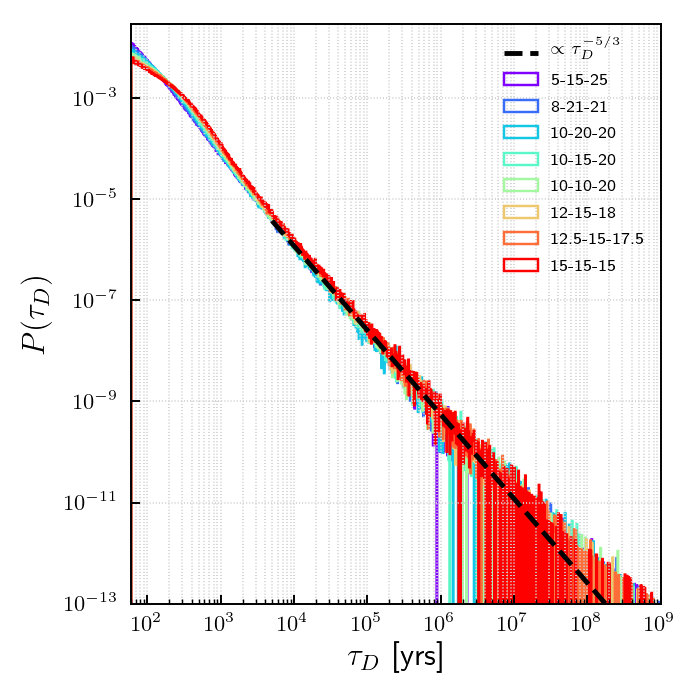}
        \end{subfigure}
        \hfill
        \begin{subfigure}[b]{0.485\textwidth}  
            \centering 
            \includegraphics[width=\textwidth]{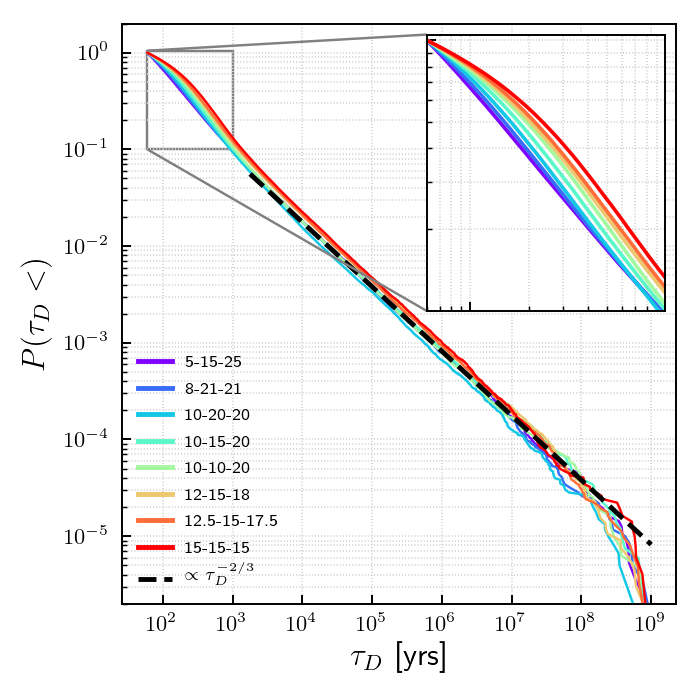}
        \end{subfigure}
        
\caption{\textit{(Left)} The differential lifetime distribution for $\tau_{D} > 60$yrs  for all three-body systems under consideration. A $-5/3$ power-law is shown for comparison with the lifetime distribution tail. \textit{(Right)} The cumulative lifetime distribution for $\tau_{D} > 60$yrs for all the three-body systems under consideration. A $-2/3$ power-law is shown for comparison with the lifetime distribution tail.}
\label{fig:lifes_dist}
\end{figure*}

\begin{figure*}
	\includegraphics[width=\textwidth]{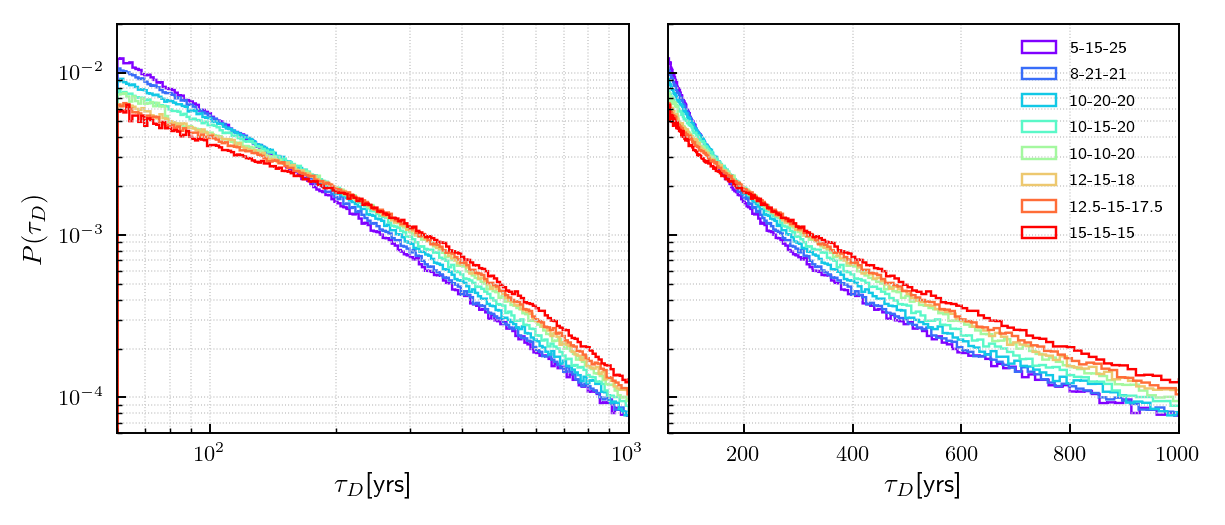}
    \caption{(\textit{Left}) Zoomed-in differential lifetime distribution for all three-body systems under consideration in log-log scale. \textit{Right} Zoomed-in differential lifetime distribution for all three-body systems under consideration in a semi-log plot.     }
    \label{fig:short_lifes}
\end{figure*}

\begin{figure*}
	\includegraphics[width=\textwidth]{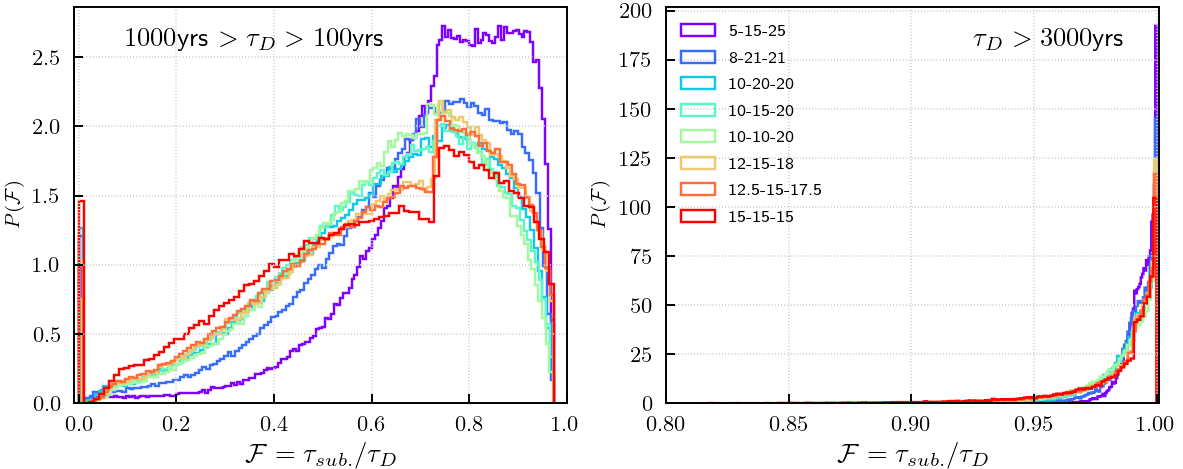}
    \caption{ The $\mathcal{F}$ quantity is the ratio of the total time spent in sub-escapes to the total lifetime of the interaction. (\textit{Left}) The distribution of $\mathcal{F}$ for $1000\text{yrs } > \tau_D > 100$yrs for all three-body systems under consideration. (\textit{Right}) The distribution of $\mathcal{F}$ for $1000\text{yrs } > \tau_D > 3000$yrs for all three-body systems under consideration. Note the different x-axis range for the 2 plots.}
    \label{fig:levy_dist}
\end{figure*}

\begin{table}
	\centering
	\caption{Differential lifetime distribution power-law index $p$ and corresponding 2$\sigma$ errors for $3\times10^{3}\text{yrs} < \tau_D < 10^{6}\text{yrs}$. The theoretical prediction is $p = -5/3 = -1.66$. }
	\label{tab:power_laws}
	\begin{tabular}{cc} 
		\hline
		Masses($M_{\odot}$) & $p$\\
		\hline
		15,15,15 & $-1.711 \pm 0.030$ \\
		12.5,15,17.5 & $-1.712 \pm 0.030$  \\
		12,15,18 & $-1.692 \pm 0.030$ \\
		10,10,20 & $-1.709 \pm 0.031$ \\
		10,15,20 & $-1.700 \pm 0.031$ \\
		10,20,20 & $-1.726 \pm 0.030$\\
		8,21,21 & $-1.711 \pm 0.031$ \\
		5,15,25 & $-1.703 \pm 0.031$ \\
		\hline
	\end{tabular}
\end{table}

We expect the differential lifetime distribution of three-body systems to follow a $-5/3$ power-law at late times as it is dominated by sub-escape motion. While at earlier times, we expect to see an exponential distribution due to ergodic motion. 

Figure~\ref{fig:lifes_dist} shows the differential (left) and cumulative (right) lifetime distributions for all the three-body systems under consideration. The black dashed line shows the predicted power-law for comparison. Table~\ref{tab:power_laws} shows the fitted power law index and the corresponding uncertainty. These are in excellent agreement with the theoretical prediction of $-5/3$. It is fascinating to note that the back-of-the-envelope calculation presented in Section~\ref{pert_pred} can accurately capture the behavior at late lifetimes. 

In Figure~\ref{fig:lifes_dist}, we notice that the power-law starts approximately at $\tau_D > 3000$yrs. For shorter lifetimes $\tau_D < 3000$yrs, we notice that the power-law no longer exists. Instead, we see a ``bump-like'' feature that is characteristic of exponential distributions in log-log scale. If the distribution is indeed exponential, then it would appear linear in a semi-log scale plot. The right panel of Figure~\ref{fig:short_lifes} shows a zoomed-in version of the differential lifetime distribution in a semi-log scale plot. The lifetime distributions do not appear linear in the semi-log scale plot indicating that the distribution is not purely exponential. However, looking at the left panel of Figure~\ref{fig:short_lifes} which shows the log-log scale version, the distribution does not appear to be a pure power-law as well because it is not linear. This indicates that on shorter time scales, one does not find only ergodic interactions. There is a \textit{mixing} phenomenon between the ergodic motion and the sub-escape motion resulting in the distribution not being purely exponential. 

This \textit{mixing} can be seen more clearly in Figure~\ref{fig:levy_dist} where the distribution of the quantity $\mathcal{F}$ is plotted. $\mathcal{F}$ is defined as 
\be 
\mathcal{F} = \frac{\tau_{sub.}}{\tau_D}
\ee
where $\tau_{sub.}$ is the total interaction time spent in sub-escapes and $\tau_D$ is the total duration/lifetime of the interaction. Therefore, $\mathcal{F}$ denotes the fraction of total time spent by an interaction in sub-escapes. Thus, if $\mathcal{F} << 1$, then a dominant fraction of the three-body motion was spent in scramble states\footnote{Prompt ejections have $\mathcal{F} = 0$, however, we are ignoring that case here.}. While, if $\mathcal{F} \sim 1$, then a dominant fraction of the three-body motion was spent in sub-escapes. Figure~\ref{fig:levy_dist} shows the distribution of $\mathcal{F}$ for different lifetime regimes of the lifetime distribution. Namely the early and late lifetime regime defined as $1000\text{yrs } > \tau_D > 100$yrs and $\tau_D>3000$yrs respectively. For the late lifetime regime in Figure~\ref{fig:levy_dist}, we see that all the interactions have $0.95 < \mathcal{F}$ with the distribution peaking strongly at $\mathcal{F} = 1$. This agrees with the expectation that interactions in the late lifetime regime are dominated by long sub-escapes. On the other hand, for the early lifetime regime, we see a interactions distributed everywhere between $0 < \mathcal{F} < 1$ along with a peak at $\mathcal{F} = 0$.The $\mathcal{F} = 0$ interactions are the ergodic interactions as they spend their entire time in a scramble state with no sub-escapes. Thus, in addition to this ergodic motion, we see non-ergodic motion as well ($\mathcal{F} > 0$). This is the \textit{mixing} phenomenon between ergodic and non-ergodic motion that results in the early lifetime distribution not being purely exponential. \citet{manwadkar20} separated out the purely ergodic interactions and showed that those do indeed follow an exponential distribution.  
 
Another interesting feature of the lifetime distributions is the increasing suppression of the ``bump'' feature or the exponential component of the distribution as the mass contrast increases, clearly seen in the right panel of Figure~\ref{fig:lifes_dist} and the left panel of Figure~\ref{fig:short_lifes}. As we go from the equal mass system (15,15,15\Mdot) to the highest mass contrast system (5,15,25\Mdot), the early lifetime distribution tends towards a pure power-law. This decreasing presence of exponential decay due to ergodic motion in high mass contrast systems is expected. An interaction exhibiting ergodic motion involves a series of scramble states where the 3 masses are in approximate energy equipartition. It is easiest for equal mass bodies to exist in equipartition, hence it has the most prominent presence of an exponential decay. However, as the mass contrast increases, i.e. the distribution of masses becomes less democratic, equipartition is difficult to achieve as the heaviest mass will tend to sit at the center of mass. Hence, scramble states in such systems are rarer and also short-lived. Thus, the presence of ergodic motion decreases as we increase the mass-contrast of the three-body system.  

\section[Analytical Model for Absorptivity]{Analytical Model for absorptivity $\calE$}
\label{analyical_model}

\begin{figure}
	\includegraphics[width=\columnwidth]{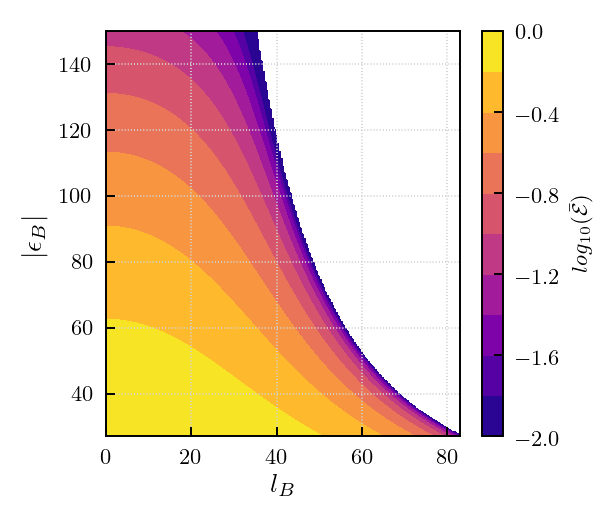}
    \caption{The contour plot for the analytical model for absorptivity $\calE$ presented in Section~\ref{ssec: comp_calE} in Equation~\ref{eqn:cale_model}. The model parameters are $p_1  = p_2 = 2$ and $E_1 = 200$ for the 15,15,15\Mdot system. It is interesting to see how this simple model can produce a contour plot similar to those seen simulations in Figure~\ref{fig:cale_log}.}
    \label{fig:cale_model}
\end{figure}

Based on the discussion in Section~\ref{ssec: comp_calE}, we construct a crude model for bi-variate averaged absorptivity $\bar{\calE}$. The model is 
\be 
\bar{\calE}\left( \eps_B, l_B \right) = \left( \frac{  (\eps_B - E_1)_{+}}{E - E_1} \right)^{p_1}  \left[ 1 - \left(\frac{l_B}{l_s(\eps_B)}\right)^{p_2} \right]
\label{eqn:cale_model}
\ee
where $p_1, p_2 > 0$, $E_1$ is the energy beyond which absorptivity vanishes, $l_s(\eps_B)$ is the maximum binary angular momentum for a given binary energy $e_B$ and is given by
\be
l_s = \sqrt{\frac{k_s}{-2\eps_B}}
\ee
The above equation is derived from Equation~\ref{binary_region}. A contour plot for this analytical model is shown in Figure~\ref{fig:cale_model} for $p_1 = p_2 = 2$ and $E_1 = 200$.  
Even though it is a crude and simplistic model, it captures all the behavior of absorptivity $\calE$ discussed in Section~\ref{ssec: comp_calE}: (i) for tight enough binaries, that is $\eps_B \leq E_1$, the absorptivity vanishes, (ii) as the binary energy approaches its maximum value at $\eps_B = E$, absorptivity $\calE$ approaches its maximum value at 1. (iii) In addition to accounting for the $\eps_B$ dependence, the model accounts for the $l_B$ dependence seen in simulations. 

To test this model and accurately probe the dependence of absorptivity on parameters like mass ratios, binary energy, binary angular momentum etc., a large number of binary-single scattering experiments will need to performed.

\section{Summary and Conclusion} 
\label{concl}

The flux-based theory exactly reduces the study of the outcome distribution to the study of the emissivity function as a function of its variables, namely $\calE (u)$. So far, we do not have a mature model for $\calE (u)$. Hence, the flux-based theory can make predictions either to $\calE (u)$-independent quantities (the critical exponent at threshold for marginal escape) or to quantities which are $\calE (u)$-blind (a posteriori, the escape probability per body).

In this work, we simulate a large number of equal mass and unequal mass non-hierarchical three-body systems to compare the statistical properties obtained from simulations to the theoretical predictions of \citet{Kol2020}. We focus on 4 statistical properties of three-body systems: (i) the ergodic escape probabilities, (ii) the threshold distribution of effective angular momentum $l_F$, (iii) the absorptivity $\calE$ and (iv) the lifetime distribution. The results and conclusions for each of these statistical comparisons are detailed below:

\begin{enumerate}
  \item The ergodic escape probabilities are in excellent agreement, down to the 1\% level, with the theoretical prediction of $P_s \propto k_s^{3/2}$. There is a big leap in accuracy in predicting escape probabilities from the \citet{Kol2020} formalism, compared to the \citet{Valtonen_book_2006}, \citet{stone19} and \citet{ginat20} formalisms.
  
  \item \citet{Kol2020} predicts a threshold distribution for $l_F$ given by Equation~\ref{lF_distrib1} with a critical exponent of $\alpha = 2$ and a minimum threshold value of $l_{F,c}$. We perform fitting on simulated distributions to obtain the best fitting value for $\alpha$ and the threshold $l_{F,c}$. We find a best fitting value of $\alpha = 1.97^{+0.42}_{-0.29}$ (median) and $\alpha = 2.06$ (mean) which is in agreement with the theoretical prediction. Excellent agreement is also seen in the threshold value where $1.02^{+0.16}_{-0.04}$ (median) and $1.06$ (mean) are the values of the fitted threshold value normalized with respect to the \citet{Kol2020} theoretical prediction.
  
  \item \citet{Kol2020} predicts the expression for absorptivity $\calE$ given by Equation~\ref{calE_pred}. Using this, we construct contour plots in $\eps_B-l_B$ space for $\calE_s / \left< \calE_s\right>$, i.e. absorptivity normalized by its global average. The calculated values for $\calE_s / \left< \calE_s\right>$ are found to be between a minimum of zero and a maximum of $\approx 2-2.5$, which is in agreement with the theoretical prediction. The observed dependence of $\calE$ on $\eps_B, l_B$ is in agreement with our theoretical understanding of binary-single scattering. 
  
  \item There is a theoretical prediction of a $-5/3$ power-law tail and an exponential decay due to ergodic motion in the late and early lifetime regimes of the lifetime distribution. There is excellent agreement between the lifetime distribution power-law tail seen in simulations and the theoretical predictions. We find that the initial part of the distribution is not purely exponential due to a \textit{mixing} phenomenon between ergodic and non-ergodic motion.  
  
\end{enumerate}

In conclusion, we find that the three-body formalism in \citet{Kol2020} is in excellent agreement with numerous statistical properties inferred from numerical simulations. Being arguably the most accurate three-body statistical theory to date, this theory has tremendous potential for providing a complete, accurate statistical description of the three-body problem. 

One may wonder about the implications of the flux-based theory to further statistical observables, such as the multi-variate outcome distribution. At present, since a mature model of $\calE (u)$ is not available, the theory does not make such predictions. In this respect, it cannot be compared with other, more developed approaches such as \cite{stone19,ginat20} (from the flux-based perspective, these approaches suggest specific expressions for $\calE (u)$), but hopefully that could be done in the near future.

\section{Future Directions}
\label{future}
The statistical comparisons presented here will serve as a foundation for our future work where we will compare more aspects of theory and simulations. A direct measurement of absorptivity $\calE$ through extensive binary-single scattering simulations will be crucial for future studies. This is because the study of the phase space volume has essentially been reduced to a study of the absorptivity. Pinning down the analytical form for absorptivity will be important for other predictions like the $\eps_B$,$l_B$ or eccentricity distribution that will be undertaken in a future work. 

Furthermore, in this work, we do not discuss in detail the analytic form of the lifetime distribution and the \textit{mixing} phenomenon seen at smaller lifetime scales. We defer a more detailed analysis of the lifetime distributions and mixing between ergodic and non-ergodic motion to a future work as well.

\section*{Acknowledgements}

This work was carried out during the COVID-19 pandemic. The authors are thankful to all the health care workers for fighting in the front-lines of this crisis. The authors would like to thank Nicholas C. Stone for discussions and valuable comments. V.M. is thankful to Allegra Hatem, Dimitriy Leksanov, Jonathan WuWong, Marie Kim, William Cerny and Yunchong Zhang for moral support while the research was conducted. N.W.C.L. gratefully acknowledges support from the Chilean government via Fondecyt Iniciaci\'on Grant \#11180005. The simulations were run on the CfCA Calculation Server at NAOJ. A.A.T. acknowledges support from JSPS KAKENHI Grant Numbers 17F17764 and 17H06360. Analyses presented in this paper were greatly aided by the following free software packages: \texttt{NumPy} (\citet{numpy}), \texttt{Matplotlib} (\citet{matplotlib}), \texttt{emcee} (\citet{emcee}) and \texttt{Jupyter} (\citet{jupyter}). This research has made extensive use of NASA\textquotesingle s Astrophysics Data System and arXiv.

\section*{Data Availability}

The \textsc{tsunami} code, the initial conditions and the simulation data underlying this article will be shared on reasonable request to the corresponding author. 



\bibliographystyle{mnras}
\bibliography{main} 



\appendix

\section{Derivations of lifetime distribution}
\label{appendix:lftm_drv}

The exponential distribution is a consequence of ergodicity. An important property of ergodic motion is its `memory-loss' property. That is, the only information that ergodic motion retains from the initial conditions are the conserved charges. This memory-loss property can be mathematically formulated as
\be
P(T > t+s  | T > t) = P(T > s) \text{ where } t,s \geq 0 
\ee
where $P(>t)$ is the probability that a three-body interaction lasts longer than time `$t$'. Using the definition of conditional probability, 
\be
P(A | B) = \frac{P (A \cap B)}{P(B)}
\ee
we have
\be 
P(T > t+s) = P(T > t)P(T > t)
\ee
The \textit{only} distribution that satisfies this property are \textit{exponential} distributions. Thus, an exponential lifetime distribution is a consequence of ergodic motion. 

The 5/3 power law tail at late times can be derived from Kepler's third law applied to long sub-escape excursions (This calculation was first presented in \citet{hut93}). As shown in \citet{manwadkar20}, the long-lived triple interactions are dominated by very long sub-escape excursions which are essentially non-terminal ejections with $E\sim0$ (but still negative as its bound). Keeping this in mind, the differential distributions of these lifetimes can be written as
\be
\frac{dN}{dt} = \left( \frac{dN}{dE} \right) \left( \frac{dE}{dt} \right)
\ee
where $\frac{dN}{dE}$ is the differential distribution of bound ejection energies and $\frac{dE}{dt}$ is the differential distribution of energies with respect to the lifetime $t$. By Kepler's third law, we can relate the ejection energy $E$ to the orbital time period as $E \propto t^{-2/3}$. Therefore, $\frac{dE}{dt} \propto t^{-5/3}$. Furthermore, the distribution of ejection energies $E$ is smooth at $E = 0$ and hence one can approximate $\frac{dN}{dE} \sim const$, see for example \eqref{dGamma}. Altogether, the differential distribution of lifetimes at late times becomes
\be
\frac{dN}{dt} \propto t^{-5/3}
\ee

\section{Convergence for Ergodic Cut}
\label{appendix:erg_cut}

\begin{figure*}
    \centering
        \begin{subfigure}[b]{\textwidth}
            \centering
            \includegraphics[width=0.65\textwidth]{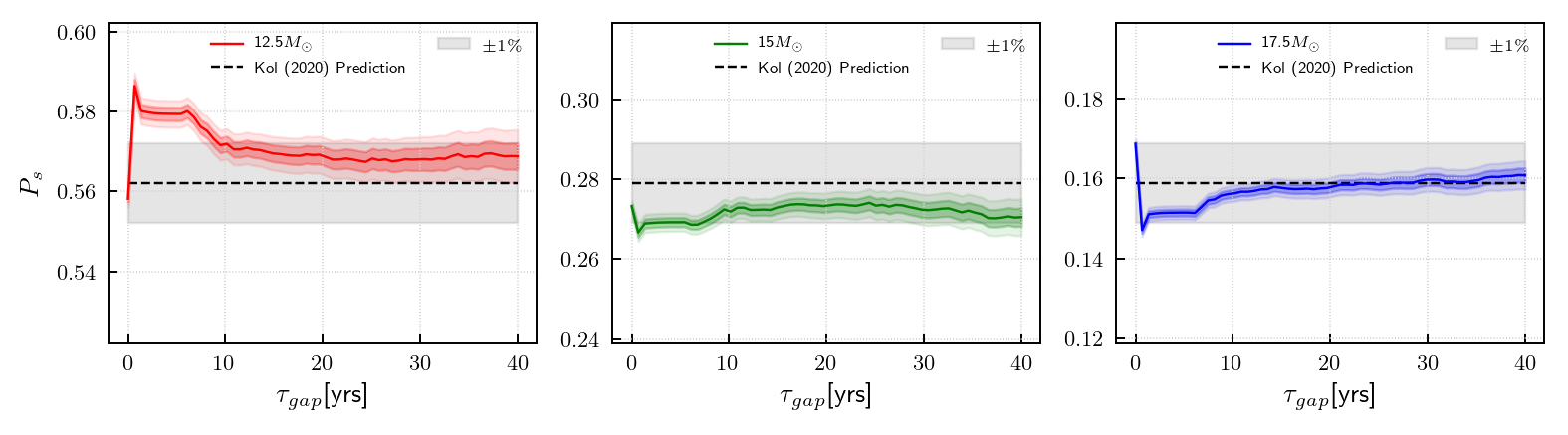}
            \caption[]%
            {{\small 12.5,15,17.5\Mdot system}}
        \end{subfigure}
        \begin{subfigure}[b]{\textwidth}
            \centering
            \includegraphics[width=0.65\textwidth]{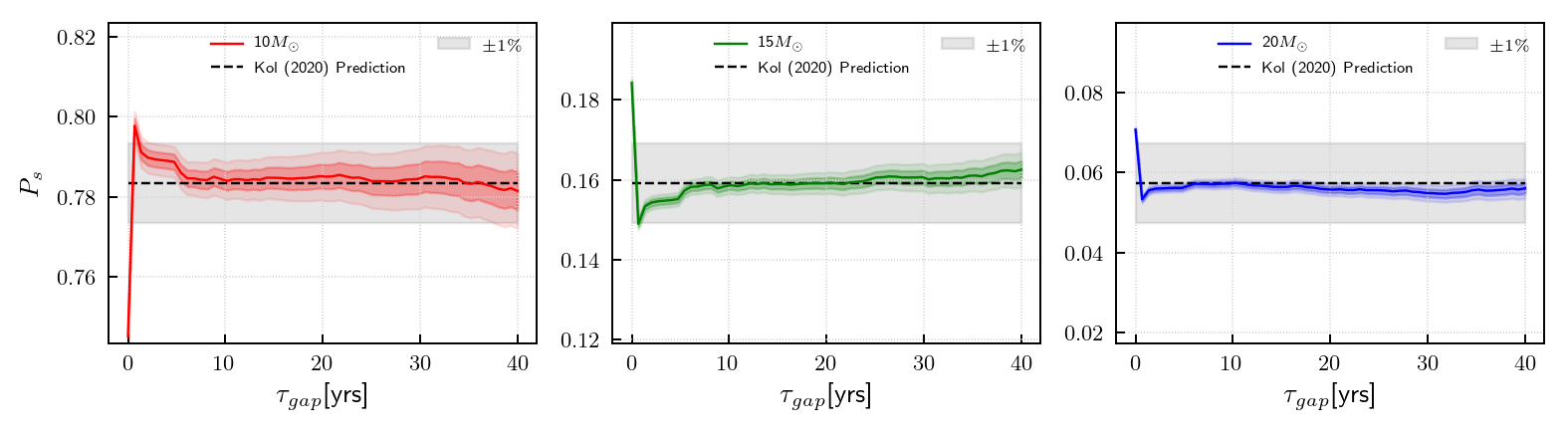}
            \caption[]%
            {{\small 10,15,20\Mdot system}}    
        \end{subfigure}
        \begin{subfigure}[b]{\textwidth}  
            \centering 
            \includegraphics[width=0.65\textwidth]{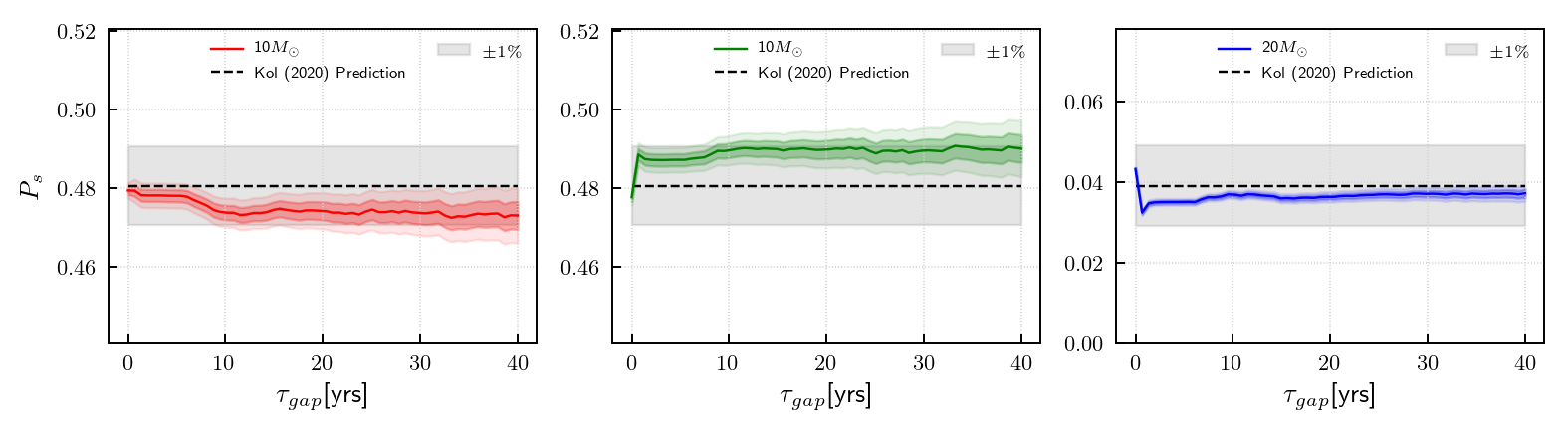}
            \caption[]%
            {{\small 10,10,20\Mdot system}}  
        \end{subfigure}
        \begin{subfigure}[b]{\textwidth}  
            \centering 
            \includegraphics[width=0.65\textwidth]{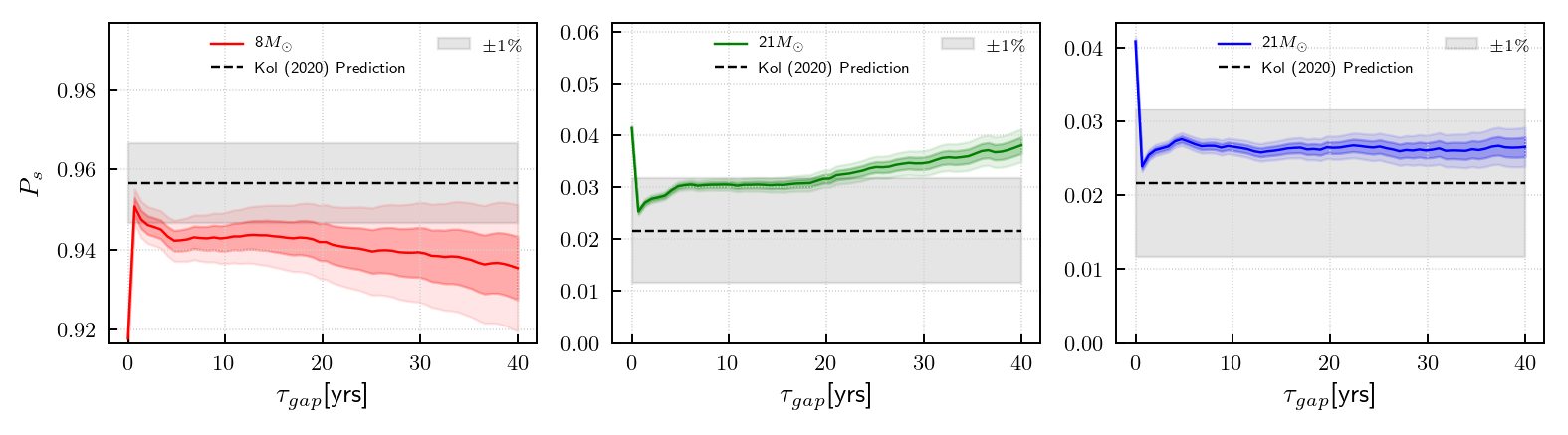}
            \caption[]%
            {{\small 8,21,21\Mdot system}}    
        \end{subfigure}
        \begin{subfigure}[b]{\textwidth}  
            \centering 
            \includegraphics[width=0.65\textwidth]{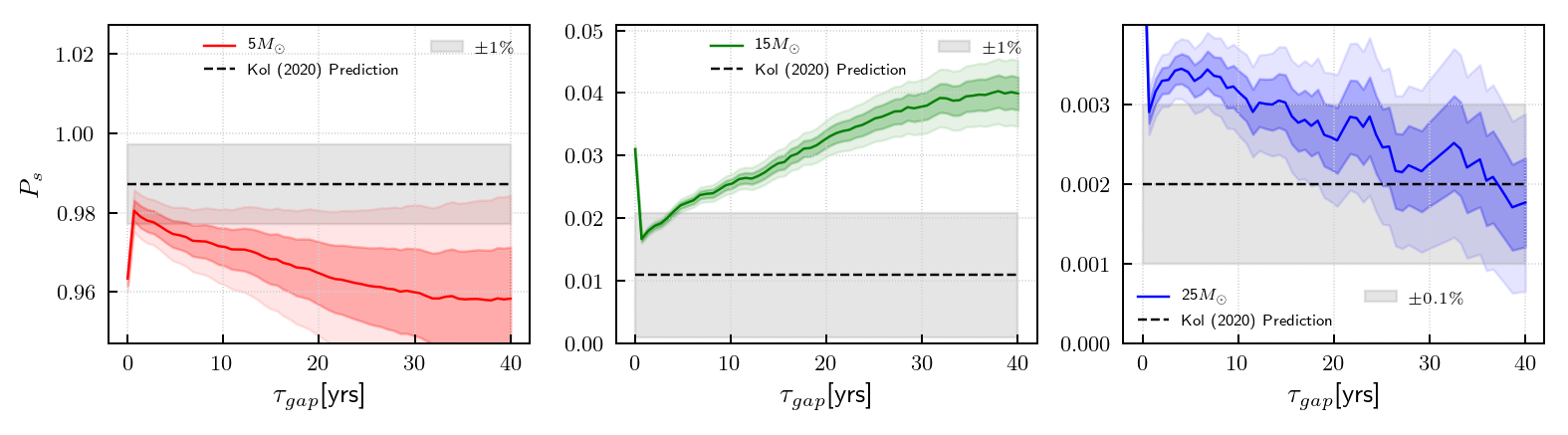}
            \caption[]%
            {{\small 5,15,25\Mdot system}}    
        \end{subfigure}

\caption{The escape probabilities $P_s$ as a function for different ergodic cuts by varying the $\tau\textsubscript{gap}$ cutoff for different three-body systems. In all the cases, the lifetime cut of $\tau_D > 80$yrs is applied. The horizontal black dashed line shows the theoretical prediction of $P_s$ value from \citet{Kol2020}. The gray shaded region is the $\pm 1\%$ region around the theoretical prediction for $P_s$. The dark- and light-colored shaded regions around the solid colored line denote the $1\sigma$ and $2\sigma$ uncertainties in ejection probabilities in simulations. The uncertainties are calculated using standard Poisson errors.}
\label{fig:appendix_taucon}
\end{figure*}

Convergence plots for the $\tau\textsubscript{gap}$ cutoff for other three-body systems are shown in Figure~\ref{fig:appendix_taucon}

\section{$L$-dependence on escape probabilities in VK06 }
\label{appendix:VK06_ps} 
According to \citet{Valtonen_book_2006}, the escape probability is given by 
\be 
P^{\text{VK06}}_{s} \propto m_s^{-q}
\ee
where $q$ is given there both a derived value $q=2$ and a fitted value, which is a function of the total angular momentum $L$ of the system and will be described below. Here we compare with the fitted value, which appears to be more accurate and hence a more challenging competition, even though given that the value of \citet{Kol2020} is derived, fairness would suggest comparing with the derived value. 

The functional form for fitted $q$ is given by 
\be 
q = \frac{3}{1 + 2\bar{L}^2}
\ee
where $\bar{L}$ is the angular momentum of three-body system normalized with respect to $L\textsubscript{max}$. Thus the index $q$ varies between $q = 3$ for $L = 0$ system and $q = 1$ for $L = L\textsubscript{max}$ system. The quantity $L\textsubscript{max}$ is the maximum angular momentum a three-body system can have for it to be bound and strongly interacting \citep[e.g.][]{Valtonen_book_2006,mikkola94}. For systems with total angular momentum $L > L\textsubscript{max}$, the systems are unbound and hierarchical. $L\textsubscript{max}$ is given by 
\be 
L\textsubscript{max} = 2.5 G \sqrt{\frac{\bar{m}^{5}}{|E|} }
\ee
where $E$ is the total energy and $\bar{m}$ is the average mass of the three-body system $\{m_1,m_2,m_3\}$ defined as 
\be 
\bar{m} = \sqrt{\frac{m_1 m_2 + m_1 m_3 + m_2 m_3}{3} }
\ee
Using this information, we can calculate the value of $q$ for the systems considered in this work. Table~\ref{tab:VK06_q} shows $L$, $L\textsubscript{max}$ and corresponding $q$ for each three-body system considered in this work. The values of $q$ are then used to calculate the predicted escape probabilities for the \citet{Valtonen_book_2006} formalism.

\begin{table}
	\centering
	\caption{The total angular momentum $L$, $L\textsubscript{max}$ and corresponding $q$ value for all three-body systems under consideration.}
	\label{tab:VK06_q}
	\begin{tabular}{cccc} 
		\hline
		Masses($M_{\odot}$) & $L$ & $L\textsubscript{max}$ & q\\
		\hline
		15,15,15 & 91.85 & 419.26 & 2.74 \\
		12.5,15,17.5 & 102.96 & 391.11 & 2.63   \\
		12,15,18 & 105.09 & 385.01 & 2.61 \\
		10,10,20 & 81.64 & 312.17 & 2.64 \\
		10,15,20 & 113.38 & 359.05 & 2.50 \\
		10,20,20 & 141.42 & 406.14 & 2.41 \\
		8,21,21 & 152.15 & 377.05 & 2.26 \\
		5,15,25 & 132.58 & 283.67 & 2.08 \\
		\hline
	\end{tabular}
\end{table}

\section[Derivation of Series formula]{Derivation of $l_F$ series formula}
\label{appendix:lf_series}

The predicted differential probability distribution for $l_F$ is given by
\be
dP_s \simeq \beta \left( l_F- l_{F,c} \right)_+^\alpha \, dl_F ,\qquad \alpha = 2
\ee
Integrating the above differential distribution, we obtain the cumulative distribution as
\be
P_s ( < l_F) = \frac{\beta}{1+\alpha} \left( l_F- l_{F,c} \right)^{\alpha+1}
\ee
Now, in our discrete simulation set, we can write
\be
P_s ( < l_F) = \frac{n}{N_T}
\ee
where $N_T$ is the size of the set under consideration and $n$ is the number of simulations where the angular momentum of the free out-going motion is less than $l_F$. Therefore,
\begin{align}
n &= N_T P_s( < l_F) \\
&= \frac{\beta N_T}{1 + \alpha} \left( l_{F,n}- l_{F,c} \right)^{\alpha+1}
\end{align}
where $l_{F,n}$ is the $n$th element in the $l_F$ series. Therefore,
\be 
\frac{n (1+\alpha)}{\beta N_T} = \left( l_{F,n}- l_{F,c} \right)^{\alpha+1}
\ee
Thus,
\begin{align}
l_{F,n} &= l_{F,c} + \left(\frac{(1+\alpha)}{\beta N_t} n \right)^{1/(1+\alpha)} \\
&= l_{F,c} + \gamma n^{1/(1+\alpha)}
\end{align}
where
\be
\gamma = \left(\frac{(1+\alpha)}{\beta N_t} \right)^{1/(1+\alpha)}
\ee
This is the derivation for the series expansion for $l_F$ that was initially given in Equation~\ref{eqn:lf_series}.

\section{Bi-variate distributions}
\label{appendix:bi_dist}

\begin{figure*}
        \centering
        \begin{subfigure}[b]{0.485\textwidth}
            \centering
            \includegraphics[width=\textwidth]{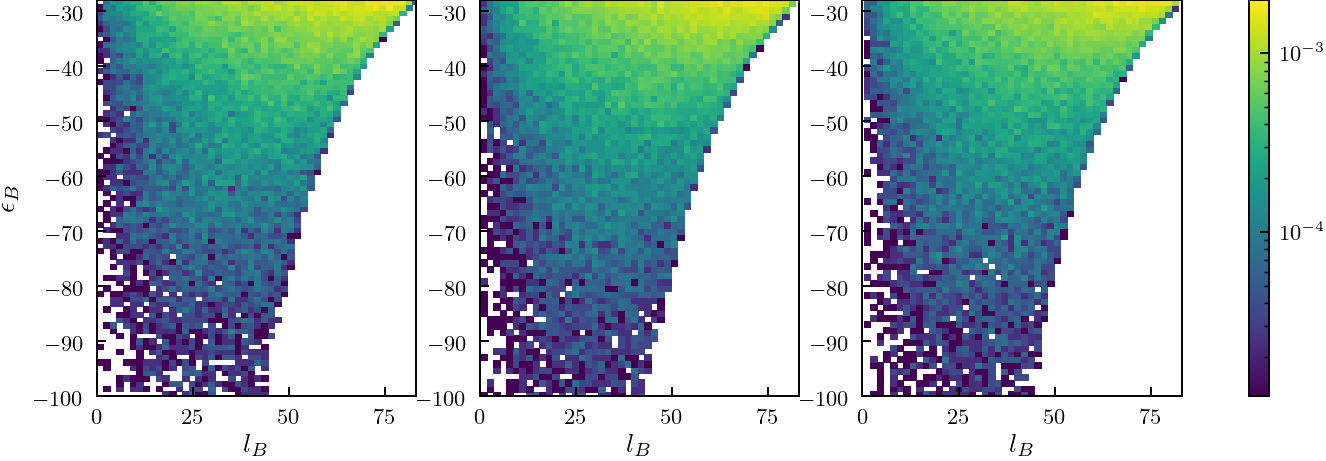}
            \caption[]%
            {{\small $e_B$-$l_B$ joint distribution for 15,15,15 system}}    
        \end{subfigure}
        \hfill
        \begin{subfigure}[b]{0.485\textwidth}  
            \centering 
            \includegraphics[width=\textwidth]{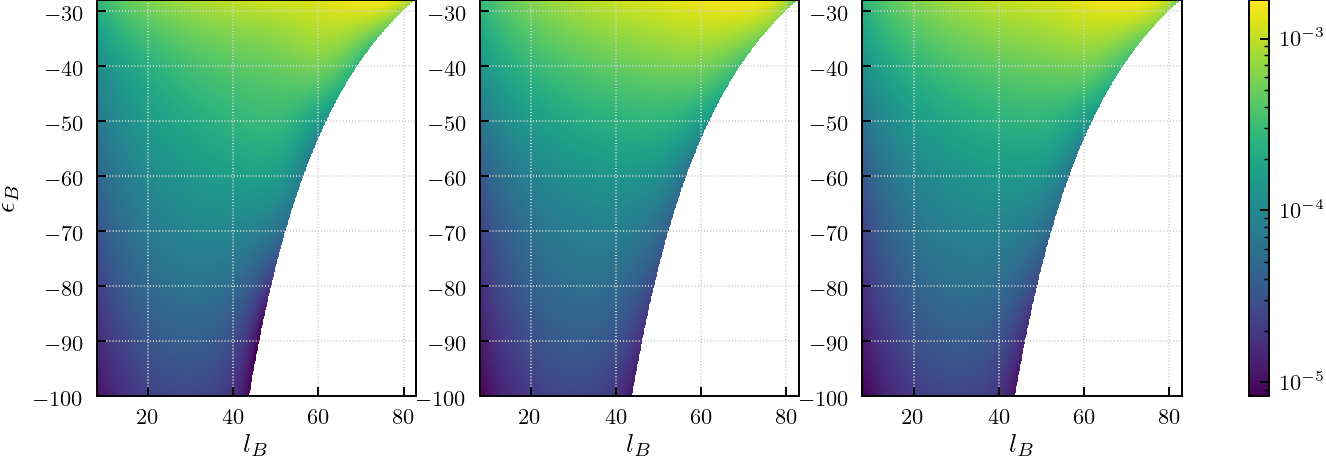}
            \caption[]%
            {{\small Smoothed $e_B$-$l_B$ joint distribution for 15,15,15 system}}    
            \label{fig:zoom_equal}
        \end{subfigure}
        \vskip\baselineskip
        \begin{subfigure}[b]{0.485\textwidth}   
            \centering 
            \includegraphics[width=\textwidth]{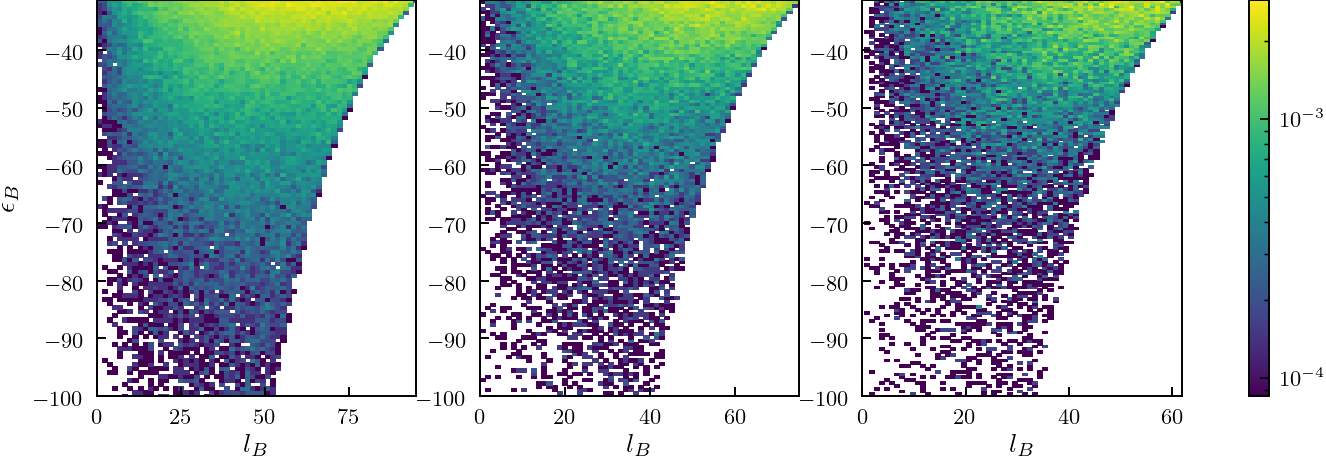}
            \caption[]%
            {{\small $e_B$-$l_B$ joint distribution for 12.5,15,17.5 system.}}    
        \end{subfigure}
        \hfill
        \begin{subfigure}[b]{0.485\textwidth}   
            \centering 
            \includegraphics[width=\textwidth]{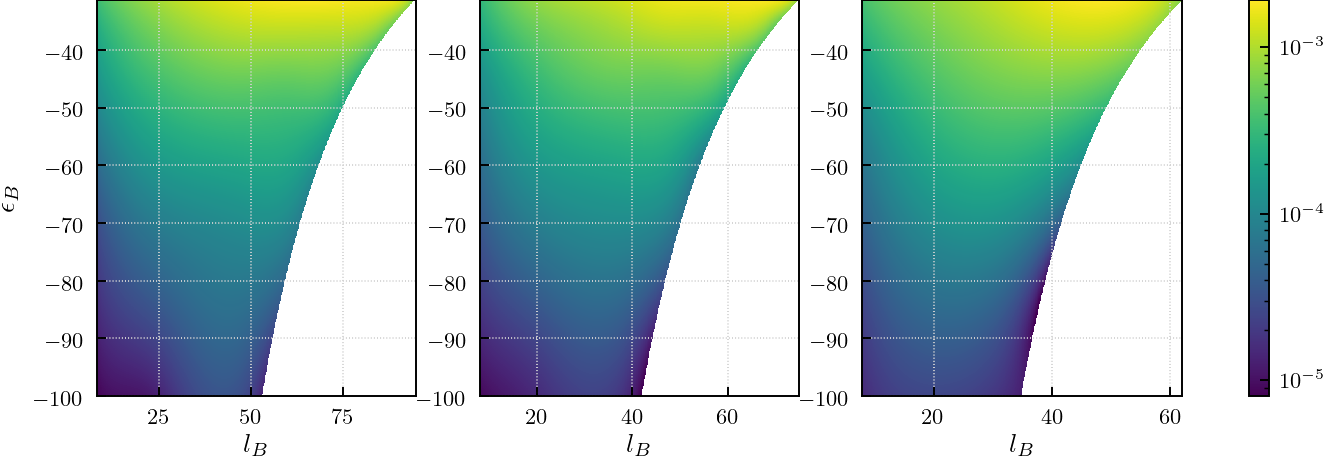}
            \caption[]%
            {{\small Smoothed $e_B$-$l_B$ joint distribution for 12.5,15,17.5 system}}    
        \end{subfigure}
        \vskip\baselineskip
        \begin{subfigure}[b]{0.485\textwidth}   
            \centering 
            \includegraphics[width=\textwidth]{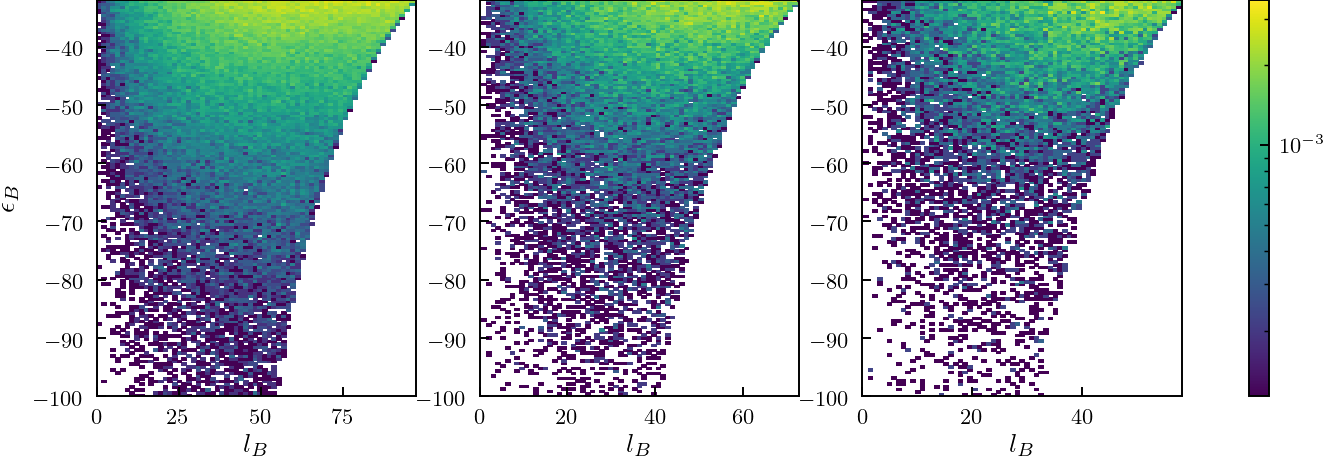}
            \caption[]%
            {{\small $e_B$-$l_B$ joint distribution for 12,15,18 system}}    
        \end{subfigure}
        \hfill
        \begin{subfigure}[b]{0.485\textwidth}   
            \centering 
            \includegraphics[width=\textwidth]{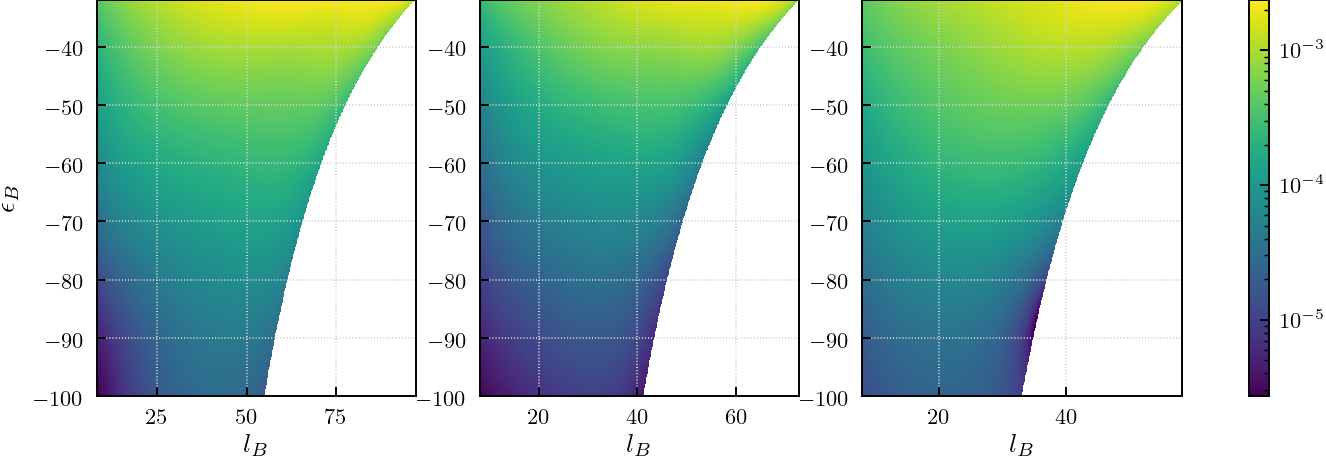}
            \caption[]%
            {{\small Smoothed $e_B$-$l_B$ joint distribution for 12,15,18 system}}    
            \label{fig:global_small}
        \end{subfigure}
        \vskip\baselineskip
        \begin{subfigure}[b]{0.485\textwidth}   
            \centering 
            \includegraphics[width=\textwidth]{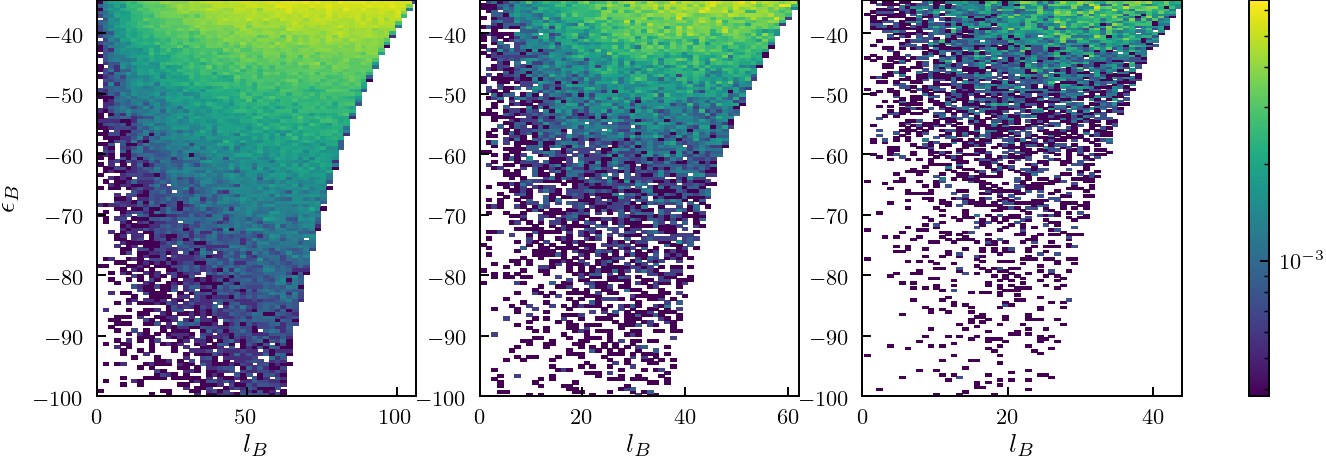}
            \caption[]%
            {{\small $e_B$-$l_B$ joint distribution for 10,15,20 system}}    
        \end{subfigure}
        \hfill
        \begin{subfigure}[b]{0.485\textwidth}   
            \centering 
            \includegraphics[width=\textwidth]{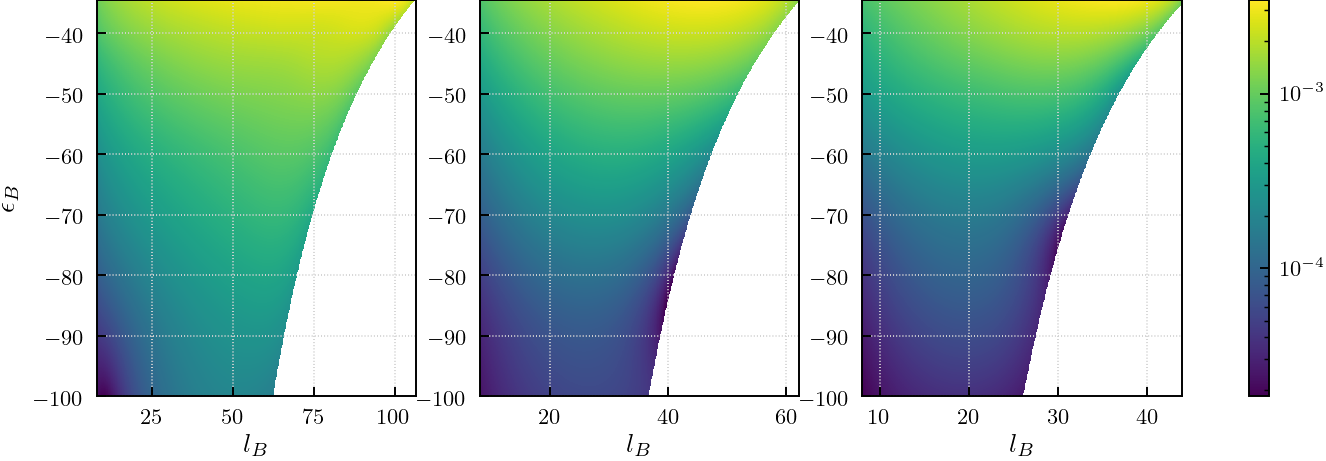}
            \caption[]%
            {{\small Smoothed $e_B$-$l_B$ joint distribution for 10,15,20 system}}    
        \end{subfigure}
\caption{The final binary energy $e_B$ and final binary angular momentum $l_B$ joint distributions for different ejection types in different three-body systems. For each set of 3 joint distributions corresponding to a single three-body system, the plots are arranged in ascending order of ejection mass from left to right.}
\label{fig:appendix_bivar}
\end{figure*}

The bi-variate distributions for final binary energy $\eps_B$ and final binary angular momentum $l_B$ are shown in Figure~\ref{fig:appendix_bivar}.

\bsp	
\label{lastpage}
\end{document}